# Matrix genetics, part 1: permutations of positions in triplets and symmetries of genetic matrices

Sergey V. Petoukhov

Department of Biomechanics, Mechanical Engineering Research Institute of the Russian Academy of Sciences
s*petoukhov@gmail.com*, *petoukhov@imash.ru*, *http://symmetry.hu/isabm/petoukhov.html*

**Abstract**: The Kronecker family of the genetic matrices is investigated, which is based on the genetic matrix [C T; A G], where C, T, A, G are the letters of the genetic alphabet. The matrix [C T; A G] in the second Kronecker power is the (4*4)-matrix of 16 duplets. The matrix [C T; A G] in the third Kronecker power is the (8*8)-matrix of 64 triplets. It is significant that peculiarities of the degeneracy of the genetic code are reflected in the symmetrical black-and-white mosaic of these genetic matrices. The article represents interesting mathematical properties of these mosaic matrices, which are connected with positional permutations inside duplets and triplets; with projector operators; with unitary matrices and cyclic groups, etc. Fractal genetic nets are proposed as a new effective tool to study long nucleotide sequences. Some results about revealing new symmetry principles of long nucleotide sequences are described.

## Contents

# 1 Introduction

An investigation of structural analogies between computer informatics and genetic informatics is one of the important tasks of modern science in a connection with a creation of DNA-computers and with a development of bioinformatics. Information is stored in computers in a form of matrices. The author considers matrix presentations of natural ensembles of elements of the genetic code. Such matrix approach reveals new structural aspects of the genetic system and proposes new possibilities to understand the question, why the nature has chosen such special system of genetic elements from huge number of other possible systems. It is important because the modern science does not know why the alphabet of genetic language has four letters (it could have any other number of letters in principle) and why just these four nitrogenous bases are chosen as elements of the genetic alphabet from billions possible chemical compounds.

This article is devoted to analysis of ensembles of molecular structures of the genetic code from the viewpoint of matrix methods of computer informatics. The used symbols A, C, G, T mean adenine, cytosine, guanine and thymine that are elements of the genetic alphabet in DNA (thymine T is replaced by uracil U in RNA). This version of the article uses the symbol T instead of the symbol U in previous versions. Also the genetic matrix [C T; A G] is used instead of the related genetic matrix [C A; U G] in previous versions (the using of the matrix [C T; A G] allows showing a connection of the genetic system with well-known types of hypercomplex numbers more easily). New essential materials are added about fractal genetic nets and Symmetry Principles of long DNA-sequences.

## 2 The natural system of binary numeration of multiplets in the genetic matrices

The Kronecker family of genomatrices $P^{(n)} = [C\ T;\ A\ G]^{(n)}$ gives complete sets of genetic multiplets in the universal mathematical form based on the square matrix of the genetic alphabet. Each genomatrix $P^{(n)}=[C\ T;\ A\ G]^{(n)}$ contains a complete set of n-plets as its matrix elements. For example, the (8*8)-genomatrix $P^{(3)}= [C\ T;\ A\ G]^{(3)}$ contains all 64 triplets, which encode 20 amino acids and stop-signals (Figure 1). The set of the letters A, C, G, T of the genetic alphabet is not an arbitrary one, but it bears an important system of binary-opposite attributes. The system of such attributes parts the four-letter alphabet into three various pairs of letters, which are equivalent from a viewpoint of one of these attributes or its absence: 1) C=T, A=G (according to binary-opposite attributes: "pyrimidine" or "purine"); 2) A=C, G=T (according to attributes: "keto" or "amino" [Waterman, 1999, § 6.3]); 3) C=G, A=T (according to attributes: three or two hydrogen bonds are materialized in these complementary pairs).

Let us number these binary-opposite attributes by numbers N = 1, 2, 3 and ascribe to each of the four letters A, C, G, T the symbol "$1_N$", if a letter has one of two attributes marked by number "N", and the symbol "$0_N$", if this letter has the opposite attribute marked by the same number N (see details on Figure 2). In that way we receive representation of the four-letter alphabet in the system of its three "binary sub-alphabets to attributes".

P =

|   | 1 | 0 |
|---|---|---|
| *1* | C<br>11(3) | T<br>10(2) |
| *0* | A<br>01(1) | G<br>00(0) |

; $P^{(2)}=$

|        | 11 (3)       | 10 (2)       | 01 (1)       | 00 (0)       |
|--------|--------------|--------------|--------------|--------------|
| *11(3)* | CC<br>1111 *(15)* | CT<br>1110 *(14)* | TC<br>1101 *(13)* | TT<br>1100 *(12)* |
| *10(2)* | CA<br>1011 *(11)* | CG<br>1010 *(10)* | TA<br>1001 *(9)* | TG<br>1000 *(8)* |
| *01(1)* | AC<br>0111 *(7)* | AT<br>0110 *(6)* | GC<br>0101 *(5)* | GT<br>0100 *(4)* |
| *00(0)* | AA<br>0011 *(3)* | AG<br>0010 *(2)* | GA<br>0001 *(1)* | GG<br>0000 *(0)* |

$P^{(3)}=$

| | **111 (7)** | **110 (6)** | **101 (5)** | **100 (4)** | **011 (3)** | **010 (2)** | **001 (1)** | **000 (0)** |
|---|---|---|---|---|---|---|---|---|
| *111* *(7)* | CCC 111111 *(63)* | CCT 111110 *(62)* | CTC 111101 *(61)* | CTT 111100 *(60)* | TCC 111011 *(59)* | TCT 111010 *(58)* | TTC 111001 *(57)* | TTT 111000 *(56)* |
| *110* *(6)* | CCA 110111 *(55)* | CCG 110110 *(54)* | CTA 110101 *(53)* | CTG 110100 *(52)* | TCA 110011 *(51)* | TCG 110010 *(50)* | TTA 110001 *(49)* | TTG 110000 *(48)* |
| *101* *(5)* | CAC 101111 *(47)* | CAT 101110 *(46)* | CGC 101101 *(45)* | CGT 101100 *(44)* | TAC 101011 *(43)* | TAT 101010 *(42)* | TGC 101001 *(41)* | TGT 101000 *(40)* |
| *100* *(4)* | CAA 100111 *(39)* | CAG 100110 *(38)* | CGA 100101 *(37)* | CGG 100100 *(36)* | TAA 100011 *(35)* | TAG 100010 *(34)* | TGA 100001 *(33)* | TGG 100000 *(32)* |
| *011* *(3)* | ACC 011111 *(31)* | ACT 011110 *(30)* | ATC 011101 *(29)* | ATT 011100 *(28)* | GCC 011011 *(27)* | GCT 011010 *(26)* | GTC 011001 *(25)* | GTT 011000 *(24)* |
| *010* *(2)* | ACA 010111 *(23)* | ACG 010110 *(22)* | ATA 010101 *(21)* | ATG 010100 *(20)* | GCA 010011 *(19)* | GCG 010010 *(18)* | GTA 010001 *(17)* | GTG 010000 *(16)* |
| *001* *(1)* | AAC 001111 *(15)* | AAT 001110 *(14)* | AGC 001101 *(13)* | AGT 001100 *(12)* | GAC 001011 *(11)* | GAT 001010 *(10)* | GGC 001001 *(9)* | GGT 001000 *(8)* |
| *000* *(0)* | AAA 000111 *(7)* | AAG 000110 *(6)* | AGA 000101 *(5)* | AGG 000100 *(4)* | GAA 000011 *(3)* | GAG 000010 *(2)* | GGA 000001 *(1)* | GGG 000000 *(0)* |

Figure 1. The beginning of the Kronecker family of symbolic genomatrices $P^{(n)}=[C\ T;\ A\ G]^{(n)}$ for n = 1, 2, 3. Each genomatrix $[C\ T;\ A\ G]^{(n)}$ has individual binary numeration for each column, each row and each n-plet due to the sub-alphabets of the genetic alphabet (see explanations in text). The decimal equivalents of these binary numbers are shown in brackets.

| | Symbols of genetic "letter" from a viewpoint of a kind of binary-opposite attributes | C | A | G | T |
|---|---|---|---|---|---|
| N=1 | $1_1$ – pyrimidine $0_1$ – purine | $1_1$ | $0_1$ | $0_1$ | $1_1$ |
| N=2 | $1_2$ – amino $0_2$ – keto | $1_2$ | $1_2$ | $0_2$ | $0_2$ |
| N=3 | $1_3$ – a letter with three hydrogen bonds $0_3$ – a letter with two hydrogen bonds | $1_3$ | $0_3$ | $1_3$ | $0_3$ |

Figure 2. Three binary sub-alphabets according to three kinds of binary-opposite attributes in the set of the nitrogenous bases C, A, G, T

The four-letter alphabet of the genetic code is curtailed into the two-letter alphabet on the basis of each kind of the attributes. For example, from the viewpoint of the first kind of binary-opposite attributes (N=1) we have the alphabet from two letters $0_1$ and $1_1$ (instead of the four-letter alphabet), which can be named "the binary sub-alphabet to first kind of binary attributes". The binary sub-alphabets of the genetic alphabet are the basis for the special system of numerations of each n-plet, each column and each row of the genomatrices $P^{(n)}$. Let us describe this system.

A binary numeration of columns and rows of the matrix $P^{(n)} = [C\ T;\ A\ G]^{(n)}$ (see Figure 1) is connected with binary symbols of letters C, A, G, T in the binary sub-alphabets №№ 1 and 2

(N=1 and N=2) respectively. More precisely, to get number of a column of the matrix [C T; A G]$^{(n)}$, one should take a sequence of letters of any n-plet from this column and write the corresponding sequence of the binary symbols of these letters from the viewpoint of the binary sub-alphabet № 1. This binary sequence is binary number of this column (all n-plets of this column are equivalent to each other from the viewpoint of binary sub-alphabet № 1). For example, let's consider the matrix [C T; A G]$^{(3)}$ and its column with a triplet CAT (Figure 1). From the viewpoint of the binary sub-alphabet № 1 (where C=T=1 and A=0), the sequence of letters CAT is equivalent to the binary sequence 101. This binary number 101 is the numeration number of the whole column because all triplets of this column have the same binary sequence 101 from the viewpoint of binary sub-alphabet № 1.

The binary number of a row of the matrix [C T; A G]$^{(n)}$ is constructed in a similar algorithmic way by interpretation of any n-plet of this row from the viewpoint of the binary sub-alphabet № 2. For example, let us consider the same triplet CAT in the matrix [C T; A G]$^{(3)}$. From the viewpoint of the binary sub-alphabet № 2 (where C=A=1 and T=0), the sequence of letters CAT is equivalent to the binary sequence _110_. This binary number _110_ is the numeration number of the matrix row with the triplet CAT because all triplets of this column have the same binary sequence _110_ from the viewpoint of binary sub-alphabet № 1.

Any matrix [C T; A G]$^{(n)}$ has a binary coordination number for each of its n-plets. All sets of its n-plets have a series of binary coordination numbers algorithmically, which are equivalent to a series of integers 0, 1, 2 ,…, ($4^n$-1) in decimal numeration system. Such coordination number of each n-plet is constructed by means of combination of binary numbers of its row and its column into the single whole in a form of 2n-digit binary number. The first half of such coordination number coincides with binary number of the matrix row of this n-plet, and the second half coincides with binary number of its column. For example, the considered triplet CAT has the individual coordinate 110101 in the matrix [C T; A G]$^{(3)}$ (Figure 1). At translation of such 2n-digit binary numbers into the decimal numeration system, all n-plets receive their individual decimal numbers from the series of integers 0, 1, 2 ,…, ($4^n$-1). All n-plets are disposed in the genomatrices regularly in a form of sequences with the monotonous change of coordinates of n-plets. For example, all triplets in the genomatrix [C T; A G]$^{(3)}$ have their natural ordering in accordance with the monotonous sequence of their coordinates 0, 1, 2,…, 63. Such natural numerations of well-ordered n-plets are useful for investigation of rules of symmetric relations among elements of various dialects of the genetic code. In this way we obtain the opportunity to work with numbers in genetic codes. In other words, we digitize elements of the genetic code.

One can replace each triplet of the genomatrix [C T; A G]$^{(3)}$ by its 6-digit binary coordinate (Figure 1). It is interesting, that such variant of the genomatrix P$^{(3)}$ is famous for a few thousand years: it is identical to the famous matrix of 64 hexagrams in Fu-Xi's order from the ancient Chinese "The Book of Changes" ("I Ching"). This matrix of 64 hexagrams was declared by ancient Chinese as the universal natural archetype [Petoukhov, 2001a,b; 2005; 2008b]. Each hexagram in the Chinese matrix is arranged by two independent trigrams, which symbolize its row and its column. But each element of the genomatrix [C T; A G]$^{(3)}$ is also connected with a binary hexagram arranged by two independent trigrams, which symbolize its row and its column. The creator of the first computer G.Leibniz was amazed by this Chinese matrix, when he became acquainted with it, because he considered himself as the originator of the binary numeration system, which was presented in this ancient matrix already. He saw in peculiarities of this matrix many materials for his ideas of a universal language and binary systems. "Leibniz has seen in this similarity the evidence of the pre-established harmony and unity of the divine plan for all epochs and for all people" [Schutskiy, 1997, c. 12]. Thereby our matrix approach to the genetic code ("matrix genetics") leads us additionally to historical analogies and to the problem of a connection of times. One can add that molecular genetics is interested for a few decades already to investigate analogies between the genetic code and the system of "I Ching" [Stent, 1969; Jacob, 1974, 1977; and others]. Modern physics pays attention to "I Ching" also [Capra, 2000].

# 3 Realignments of the genomatrix [C A; T G]$^{(3)}$ by permutations of positions in triplets

Now we continue the analysis of matrix presentations of the genetic code {see [Petoukhov, arXiv:0802.3366 (q-Bio.Qm)]}. Modern science recognizes many variants (or dialects) of the genetic code, data about which are shown on the NCBI's website http://www.ncbi.nlm.nih.gov/Taxonomy/Utils/wprintgc.cgi. 17 variants (or dialects) of the genetic code exist which differ one from another by some details of correspondences between triplets and objects encoded by them. Most of these dialects (including the so called Standard Code and the Vertebrate Mitochondrial Code) have the following general scheme of their degeneracy where 32 "black" triplets with "strong roots" and 32 "white" triplets with "weak" roots exist.

In this general or basic scheme, the set of 64 triplets contains 16 subfamilies of triplets, every one of which contains 4 triplets with the same two letters on the first positions of each triplet (an example of such subsets is the case of the four triplets CAC, CAA, CAT, CAG with the same two letters CA on their first positions). We shall name such subfamilies as the subfamilies of *NN*-triplets. In the described basic scheme of degeneracy, the set of these 16 subfamilies of *NN*-triplets is divided into two equal subsets from the viewpoint of degeneration properties of the code (Figure 3). The first subset contains 8 subfamilies of so called "two-position" *NN*-triplets, a coding value of which is independent of a letter on their third position. An example of such subfamilies is the four triplets CGC, CGA, CGT, CGC (Figure 3), all of which encode the same amino acid Arg, though they have different letters on their third postion. All members of such subfamilies of *NN*-triplets are marked by black color in Figures 3 and 4.

The second subset contains 8 subfamilies of "three-position" NN-triplets, the coding value of which depends on a letter on their third position. An example of such subfamilies in Figure 3 is the four triplets CAC, CAA, CAT, CAC, two of which (CAC, CAT) encode the amino acid His and the other two (CAA, CAG) encode another amino acid Gln. All members of such subfamilies of NN-triplets are marked by the white color in the genomatrix P$^{(3)}$ = [C T; A G]$^{(3)}$ on Figure 4. So the genomatrix [C T; A G]$^{(3)}$ has 32 black triplets and 32 white triplets. Each subfamily of four NN-triplet is disposed in an appropriate (2x2)-subquadrant of the genomatrix [C T; A G]$^{(3)}$ due to the Kronecker algorithm of construction of the genomatrix [C T; A G]$^{(3)}$ of triplets from the alphabet genomatrix P = [C T; A G] (Figure 1).

Here one should recall the work by Rumer [Rumer, 1968] where a combination of letters on the two first positions of each triplet was named as a "root" of this triplet. A set of 64 triplets contains 16 possible variants of such roots. Taking into account properties of triplets, Rumer has divided the set of 16 possible roots into two subsets with eight roots in each. Roots CC, CT, CG, AC, TC, GC, GT, GG form the first of such octets. They were named by Rumer "strong roots". The other eight roots CA, AA, AT, AG, TA, TT, TG, GA form the second octet and they were named weak roots. When Rumer published his works, the Vertebrate Mitochondrial code and some of the other code dialects were unknown. But one can check easily that the set of 32 black (white) triplets, which we show on Figure 3 for cases of the Standard code and the Vertebrate Mitochondrial Code, is identical to the set of 32 triplets with strong (weak) roots described by Rumer. So, using notions proposed by Rumer, the black triplets can be named as triplets with the strong roots and the white triplets can be named as triplets with the weak roots. Rumer believed that this symmetrical division into two binary-oppositional categories of roots is very important for understanding the nature of the genetic code systems.

One can check easily on the basis of data from the NCBI's website (http://www.ncbi.nlm.nih.gov/Taxonomy/Utils/wprintgc.cgi) that the following 11 dialects of the genetic code have the same basic scheme of degeneracy with 32 black triplets and with 32 white triplets: 1) the Standard Code; 2) the Vertebrate Mitochondrial Code; 3) the Yeast Mitochondrial Code; 4) the Mold, Protozoan, and Coelenterate Mitochondrial Code and the Mycoplasma/Spiroplasma Code; 5) the Ciliate, Dasycladacean and Hexamita Nuclear Code; 6) the Euplotid Nuclear Code; 7) the Bacterial and Plant Plastid Code; 8) the Ascidian

Mitochondrial Code; 9) the Blepharisma Nuclear Code; 10) the Thraustochytrium Mitochondrial Code; 11) the Chlorophycean Mitochondrial Code. In this article we will consider this basic scheme of the degeneracy which is presented by means of a black-and-white mosaics in a family of genetic matrices ($P^{(3)} = [C\ T;\ A\ G]^{(3)}$, etc.) on Figure 4.

One can mentioned that the other 6 dialects of the genetic code have only small differences from the described basic scheme of degeneracy: the Invertebrate Mitochondrial Code; the Echinoderm and Flatworm Mitochondrial Code; the Alternative Yeast Nuclear Code; The Alternative Flatworm Mitochondrial Code; the Trematode Mitochondrial Code; the Scenedesmus obliquus mitochondrial Code.

According to general traditions, the theory of symmetry studies initially those natural objects that possess the most symmetrical character, and then it constructs a theory for cases of violations of this symmetry in other kindred objects. For this reason one should pay special attention to the Vertebrate Mitochondrial code which is the most symmetrical code among dialects of the genetic code and which corresponds to the basic scheme of the degeneracy. One can mention additionally that some authors consider this dialect not only as the most "perfect" but also as the most ancient dialect [Frank-Kamenetskiy, 1988], but this last aspect is a debatable one. Figure 3 shows the correspondence between the set of 64 triplets and the set of 20 amino acids with stop-signals (Stop) of protein synthesis in the Standard Code and in the Vertebrate Mitochondrial Code.

| THE STANDARD CODE | |
|---|---|
| 8 subfamilies of the "two-position NN-triplets" ("black triplets") and the amino acids, which are encoded by them | 8 subfamilies of the "three-position NN-triplets" ("white triplets") and the amino acids, which are encoded by them |
| CCC, CCT, CCA, CCG ➔ Pro | CAC, CAT, CAA, CAG ➔ His, His, Gln, Gln |
| CTC, CTT, CTA, CTG ➔ Leu | AAC, AAT, AAA, AAG ➔ Asn, Asn, Lys, Lys |
| CGC, CGT, CGA, CGG ➔ Arg | ATC, ATT, ATA, ATG ➔ Ile, Ile, Ile, Met |
| ACC, ACT, ACA, ACG ➔ Thr | AGC, AGT, AGA, AGG ➔ Ser, Ser, Arg, Arg |
| TCC, TCT, TCA, TCG ➔ Ser | TAC, TAT, TAA, TAG ➔ Tyr, Tyr, Stop, Stop |
| GCC, GCT, GCA, GCG ➔ Ala | TTC, TTT, TTA, TTG ➔ Phe, Phe, Leu, Leu |
| GTC, GTT, GTA, GTG ➔ Val | TGC, TGT, TGA, TGG ➔ Cys, Cys, Stop, Trp |
| GGC, GGT, GGA, GGG ➔ Gly | GAC, GAT, GAA, GAG ➔ Asp, Asp, Glu, Glu |

| THE VERTEBRATE MITOCHONDRIAL CODE | |
|---|---|
| 8 subfamilies of the "two-position NN-triplets" ("black triplets") and the amino acids, which are encoded by them | 8 subfamilies of the "three-position NN-triplets" ("white triplets") and the amino acids, which are encoded by them |
| CCC, CCT, CCA, CCG ➔ Pro | CAC, CAT, CAA, CAG ➔ His, His, Gln, Gln |
| CTC, CTT, CTA, CCG ➔ Leu | AAC, AAT, AAA, AAG ➔ Asn, Asn, Lys, Lys |
| CGC, CGT, CGA, CGG ➔ Arg | ATC, ATT, ATA, ATG ➔ Ile, Ile, Met, Met |
| ACC, ACT, ACA, ACG ➔ Thr | AGC, AGT, AGA, AGG ➔ Ser, Ser, Stop, Stop |
| TCC, TCT, TCA, TCG ➔ Ser | TAC, TAT, TAA, TAG ➔ Tyr, Tyr, Stop, Stop |
| GCC, GCT, GCA, GCG ➔ Ala | TTC, TTT, TTA, TTG ➔ Phe, Phe, Leu, Leu |
| GTC, GTT, GTA, GTG ➔ Val | TGC, TGT, TGA, TGG ➔ Cys, Cys, Trp, Trp |
| GGC, GGT, GGA, GGG ➔ Gly | GAC, GAT, GAA, GAG ➔ Asp, Asp, Glu, Glu |

Figure 3. Two examples of the basic scheme of the genetic code degeneracy with 32 "black" triplets and 32 "white" triplets. Top: the case of the Standard Code. Bottom: the case of the Vertebrate Mitochondrial Code. Yellow color highlights the triplets which changed their code meanings in relation to the Standard Code. All initial data are taken from the NCBI's web-site *http://www.ncbi.nlm.nih.gov/Taxonomy/Utils/wprintgc.cgi*.

The author has revealed that the disposition of the black and white triplets in the genomatrix $P^{(3)} = [C\ T;\ A\ G]^{(3)}$ for this basic scheme of the genetic code degeneracy gives the very symmetrical black-and-white mosaic of the code degeneracy (Figure 4), though the most variants of their possible dispositions give quite asymmetric mosaics [Petoukhov, arXiv:0802.3366 (q-bio.QM)].

It is much unexpected that any kind of permutation of positions in triplets, which is accomplished in all 64 triplets simultaneously, leads to the transformed genomatrix, which possesses a symmetrical mosaic of degeneracy also. The six kinds of sequences of positions in triplets exist: 1-2-3, 2-3-1, 3-1-2, 1-3-2, 2-1-3, 3-2-1. It is obvious that if one changes the positional sequence 1-2-3 in triplets, for example, by the sequence 2-3-1, the most triplets change their disposition in the genomatrix. And the initial genomatrix is reconstructed cardinally into the new mosaic matrix. For instance, in the result of such permutation the black triplet CGA is replaced in its matrix cell by the white triplet GAC, etc. Let us denote the six genomatrices, which correspond to the mentioned kinds of positional sequences in triplets, by the symbols $P^{CTAG(3)}_{123}(=P^{(3)}=[C\ T;\ A\ G]^{(3)})$, $P^{CTAG(3)}_{231}$, $P^{CTAG(3)}_{312}$, $P^{CTAG(3)}_{132}$, $P^{CTAG(3)}_{213}$, $P^{CTAG(3)}_{321}$. Here the bottom indexes show the appropriate positional sequences in triplets; the upper index shows the kind of basic genomatrix [C T; A G] of the Kronecker family (later we shall consider other cases of such basic genomatrices). Figure 4 demonstrates these six genomatrices.

| CCC | CCT | CTC | CTT | TCC | TCT | TTC | TTT |
|---|---|---|---|---|---|---|---|
| CCA | CCG | CTA | CTG | TCA | TCG | TTA | TTG |
| CAC | CAT | CGC | CGT | TAC | TAT | TGC | TGT |
| CAA | CAG | CGA | CGG | TAA | TAG | TGA | TGG |
| ACC | ACT | ATC | ATT | GCC | GCT | GTC | GTT |
| ACA | ACG | ATA | ATG | GCA | GCG | GTA | GTG |
| AAC | AAT | AGC | AGT | GAC | GAT | GGC | GGT |
| AAA | AAG | AGA | AGG | GAA | GAG | GGA | GGG |

| CCC | CTC | TCC | TTC | CCT | CTT | TCT | TTT |
|---|---|---|---|---|---|---|---|
| CAC | CGC | TAC | TGC | CAT | CGT | TAT | TGT |
| ACC | ATC | GCC | GTC | ACT | ATT | GCT | GTT |
| AAC | AGC | GAC | GGC | AAT | AGT | GAT | GGT |
| CCA | CTA | TCA | TTA | CCG | CTG | TCG | TTG |
| CAA | CGA | TAA | TGA | CAG | CGG | TAG | TGG |
| ACA | ATA | GCA | GTA | ACG | ATG | GCG | GTG |
| AAA | AGA | GAA | GGA | AAG | AGG | GAG | GGG |

| CCC | CCT | TCC | TCT | CTC | CTT | TTC | TTT |
|---|---|---|---|---|---|---|---|
| CCA | CCG | TCA | TCG | CTA | CTG | TTA | TTG |
| ACC | ACT | GCC | GCT | ATC | ATT | GTC | GTT |
| ACA | ACG | GCA | GCG | ATA | ATG | GTA | GTG |
| CAC | CAT | TAC | TAT | CGC | CGT | TGC | TGT |
| CAA | CAG | TAA | TAG | CGA | CGG | TGA | TGC |
| AAC | AAT | GAC | GAT | AGC | AGT | GGC | GGT |
| AAA | AAG | GAA | GAG | AGA | AGG | GGA | GGG |

| CCC | TCC | CTC | TTC | CCT | TCT | CTT | TTT |
|---|---|---|---|---|---|---|---|
| ACC | GCC | ATC | GTC | ACT | GCT | ATT | GTT |
| CAC | TAC | CGC | TGC | CAT | TAT | CGT | TGT |
| AAC | GAC | AGC | GGC | AAT | GAT | AGT | GGT |
| CCA | TCA | CTA | TTA | CCG | TCG | CTG | TTG |
| ACA | GCA | ATA | GTA | ACG | GCG | ATG | GTG |
| CAA | TAA | CGA | TGA | CAG | TAG | CGG | TGG |
| AAA | GAA | AGA | GGA | AAG | GAG | AGG | GGG |

| CCC | TCC | CCT | TCT | CTC | TTC | CTT | TTT |
|---|---|---|---|---|---|---|---|
| ACC | GCC | ACT | GCT | ATC | GTC | ATT | GTT |
| CCA | TCA | CCG | TCG | CTA | TTA | CTG | TTG |
| ACA | GCA | ACG | GCG | ATA | GTA | ATG | GTG |
| CAC | TAC | CAT | TAT | CGC | TGC | CGT | TGT |
| AAC | GAC | AAT | GAT | AGC | GGC | AGT | GGT |
| CAA | TAA | CAG | TAG | CGA | TGA | CGG | TGG |
| AAA | GAA | AAG | GAG | AGA | GGA | AGG | GGG |

| CCC | CTC | CCT | CTT | TCC | TTC | TCT | TTT |
|-----|-----|-----|-----|-----|-----|-----|-----|
| CAC | CGC | CAT | CGT | TAC | TGC | TAT | TGT |
| CCA | CTA | CCG | CTG | TCA | TTA | TCG | TTG |
| CAA | CGA | CAG | CGG | TAA | TGA | TAG | TGG |
| ACC | ATC | ACT | ATT | GCC | GTC | GCT | GTT |
| AAC | AGC | AAT | AGT | GAC | GGC | GAT | GGT |
| ACA | ATA | ACG | ATG | GCA | GTA | GCG | GTG |
| AAA | AGA | AAG | AGG | GAA | GGA | GAG | GGG |

Figure 4. The genomatrices $P^{CTAG}_{123}{}^{(3)}$ (or $P^{(3)}$ in Figure 1), $P^{CTAG}_{231}{}^{(3)}$, $P^{CTAG}_{213}{}^{(3)}$, $P^{CTAG}_{321}{}^{(3)}$, $P^{CTAG}_{312}{}^{(3)}$, $P^{CTAG}_{132}{}^{(3)}$. Each matrix cell has a triplet and an amino acid (or stop-signal) coded by this triplet. The black-and-white mosaic presents a specificity of the basic scheme of the genetic code degeneracy.

It should be mentioned that the quantity of variants of possible dispositions of 64 genetic triplets in 64 cells of the genomatrix $[C\ T;\ A\ G]^{(3)}$ is equal to the huge number $64! \sim 10^{89}$; the most of these variants have not symmetries in a disposition of these black triplets and white triplets in (8*8)-matrix, of course. But each of these six genomatrices has a symmetric character unexpectedly (Figure 4). For example, the first genomatrix $P^{CTAG}_{123}{}^{(3)} = [C\ T;\ A\ G]^{(3)}$ has the following symmetric features:
1. The upper and lower halves of this matrix are mirror-antisymmetric to each other by its colors: any pair of cells, disposed by the mirror-symmetric manner in these halves, has opposite colors.
2. Diagonal quadrants of the matrix are identical to each other from the viewpoint of their mosaics.
3. The adjacent columns 0-1, 2-3, 4-5, 6-7 are identical to each other from the viewpoint of the mosaic and of the disposition of the same amino acids in their proper cells.
4. Mosaics of all columns of the (8x8)-genomatrix and of its (4x4)-quadrants have a meander-line character, which is connected with Rademacher functions from theory of digital signal processing.
5. The turning of the genomatrix $P^{CTAG}_{123}{}^{(3)}$ into a cylinder with an agglutination of its upper and lower borders reveals a symmetric pattern of cyclic shifts. This pattern is demonstrated more clearly by the tessellation of a plane with this mosaic genomatrix $P^{CTAG}_{123}{}^{(3)}$ (Figure 5, left).

Each of the other 5 genomatrices on Figure 4 has symmetrical properties also, first of all, the following two properties:
1. The upper and lower halves of each matrix are mirror-antisymmetric to each other by its colors:any pair of cells, disposed by the mirror-symmetric manner in these halves, has opposite colors.
2. Mosaics of all columns of each (8x8)-genomatrix and of its (4x4)-quadrants have a meander-line character, which is connected with Rademacher functions from theory of digital signal processing.

## 4 The tessellation of a plane by the mosaics of the genetic matrices

The plane with the tessellation by the mosaic genomatrix $P^{CTAG}_{123}{}^{(3)} = [C\ T;\ A\ G]^{(3)}$ has the ornamental pattern with two pattern units. These two pattern units are identical in their forms, but they are inverse in their colors (black and white) and orientations (left and right). This pattern has the character of cyclic shifts that permits to think about a possible genetic meaning of cyclic codes, which play significant role in the theory of digital signal processing.

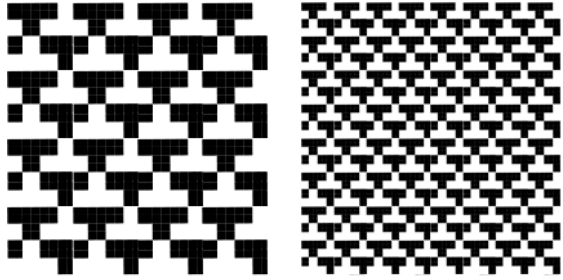

Figure 5. At the left: the tessellation of a plane by the mosaics of $P^{CTAG}_{123}{}^{(3)}$ from Figure 4.
At the right: the tessellation of a plane by the mosaics of $P^{CTAG}_{213}{}^{(3)}$ from Figure 4.

The second genomatrix $P^{CAUG}_{231}{}^{(3)}$ on Figure 4 is generated by the cyclic shift of positions in triplets (2-3-1 instead of 1-2-3). This genomatrix has the symmetric features also:
1. The upper and lower halves of this matrix are mirror-antisymmetric to each other by their colors.
2. All (4x4)-quadrants are identical to each other by their mosaics.
3. The left and the right halves of the genomatrix are identical to each other from the viewpoint of dispositions of all amino acids and stop-signals in the case of the vertebrate mitochondrial code from Fig.3.
4. The genomatrix has 4 pairs of identical columns again: 0-4, 1-5, 2-6, 3-7 that are not adjacent columns in this matrix.
5. All columns of the (8x8)-genomatrix and its (4x4)-quadrants have a meander-line character again, which is connected with Rademacher functions.

Note, that the mosaic of the initial (8x8)-genomatrix $P^{CTAG}_{123}{}^{(3)}$ is reproduced in (4x4)-quadrants of this $P^{CTAG}_{213}{}^{(3)}$ in a fractal manner: the coefficient of the fractal ranging of areas is equal to 4. The tessellations of a plane by the mosaics of $P^{CTAG}_{123}{}^{(3)}$ and of $P^{CTAG}_{213}{}^{(3)}$ demonstrate their fractal correspondence very clearly (Figure5). Such scale transformation of areas in the mosaics of the code degeneration will be named the "tetra-reproduction" transformation. The cyclic-generated genomatrix $P^{CTAG(3)}_{231}$ has the quantity of the pattern units 4 times more than the initial genomatrix $P^{CTAG(3)}_{123}$ due to this tetra-reproduction (Figures 4 and 5). This fact is interesting because an analogical tetra-reproduction (or a tetra-division) exists in the living nature always in a course of division of gametal cells, which are transmitters of genetic information. In this mysterious act of meiosis, one gamete is divided into four new gametes (this fact was mentioned specially by Schrodinger in his famous book [Schrodinger, 1955, §13]). The described tetra-reproduction of the mosaics of the genomatrices can be utilized, in particular, in formal models of meiosis. We will return to this reproduction property below in Section 9.

Presented materials of the matrix genetics lead us to questions of biological meaning. Really, we revealed unexpectedly that a simple algorithmic re-packing (re-arrangement) of elements in triplets by the cyclic shift is sufficient to receive new genomatrix with the fractal tetra-reproducing of mosaics of the code degeneration. It seems that a similar re-packing of molecular elements in biological object can be sufficient also to provide foundations of a process of tetra-reproducing in some cases, first of all, in the case of meiosis. These and other considerations permit us to put forward a hypothesis of molecular re-packing. According to this hypothesis, the mysterious process of meiosis is based on a mechanism of algorithmic re-arrangement of molecular elements of gametes with a participation of algorithms of cyclic and dyadic shifts. In our opinion, the principle of re-packing of biological molecules and of their ensembles is an important general principle of biological self-organization. It is interesting also that one can compare the tetra-division of material gametes with the tetra-division of the code genomatrices, which are information objects. These materials testify that meiosis is not an accidental material process but it is coordinated with more ancient information structures of the genetic code in their matrix form.

Now let us additionally consider on Figure 4 the genomatrix $P^{CTAG}_{321}{}^{(3)}$ with the inverse order of positions in all triplets (3-2-1 instead of 1-2-3). One can compare its mosaic with the mosaic of the $P^{CTAG}_{213}{}^{(3)}$ based on the cyclic shift of positions in all triplets: 2-1-3 instead of 3-2-1. In this case the similar phenomenon of the tetra-reproduction of these mosaics becomes apparent again but with new pattern (Figure 6).

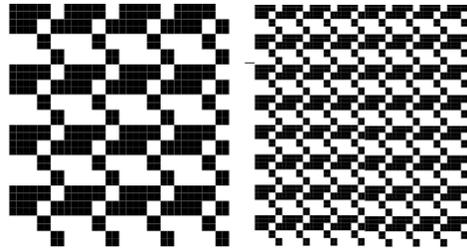

Figure 6. The tessellations of a plane by the mosaics of the genomatrices $P^{CTAG}_{213}$ (at the left) and $P^{CTAG}_{321}$ (on the right)

One can note also that all six genomatrices on Figure 4 are interconnected by special permutations of their columns and rows. The same genomatrices can be received from the initial genomatrix $P^{CTAG}_{123}{}^{(3)}$ by appropriate permutations of positions in binary 3-digit numeration of their columns and rows. In other words, the "local" permutations of positions in triplets give the same results as "global" permutations of positions in binary 3-digit numeration of columns and of rows. All six genomatrices on Figure 4 are connected with Hadamard matrices on the basis of the T-algorithm (its another name is the U-algorithm) described in [Petoukhov, 2005; 2008 a, b]. The presented permutations gave interesting results in their application to genomatrices. It seems that applications of similar permutations to genetic sequences of triplets can give interesting results also. Our study of properties of permutations in structural regularities of long nucleotide sequences is briefly described in the end of the article in a connection with fractal genetic nets.

**5 The genetic encoding and a fast Fourier transformation**

The revelation of the permutation group of the six symmetric genomatrices $P^{CTAG}_{123}{}^{(3)}$ (or $P^{(3)}$ in Figure 1), $P^{CTAG}_{231}{}^{(3)}$, $P^{CTAG}_{213}{}^{(3)}$, $P^{CTAG}_{321}{}^{(3)}$, $P^{CTAG}_{312}{}^{(3)}$, $P^{CTAG}_{132}{}^{(3)}$ seems to be the essential fact additionally because of heuristic associations with the mathematical theory of digital signal processing, where similar permutations are utilized for a long time as the useful tool. For example, the book [Ahmed, Rao, 1975, § 4.6] gives the example of the important role of the method of data permutation and of binary inversion for one of variants of the algorithm of a fast Fourier transformation. In this example the numeric sequence 0, 1, 2, 3, 4, 5, 6, 7 is reformed into the sequence 0, 4, 2, 6, 1, 5, 3, 7. But the same change of the numeration of the columns and the rows takes a place in our case (Figure 4) where the genomatrix $P^{CTAG}_{123}$ is reformed into the genomatrix $P^{CTAG}_{321}$ in the result of the inversion of the binary numeration of the columns and the rows (or of the inversion of the positions in the triplets). These and other facts permit to think that the genetic system has a connection with a fast Fourier transformation (or with a fast Hadamard transformation) [Petoukhov, 2006, 2008b].

**6 The tetra-reproduction of the genomatrices in their binary presentation**

Why the nature has chosen this variant of degeneration of genetic code, which gives such mosaics? Whether these six "triplets-permutations" genomatrices $P^{CTAG}_{123}{}^{(3)}$, $P^{CTAG}_{231}{}^{(3)}$, $P^{CTAG}_{213}{}^{(3)}$, $P^{CTAG}_{321}{}^{(3)}$, $P^{CTAG}_{312}{}^{(3)}$, $P^{CTAG}_{132}{}^{(3)}$ have such mutual mathematical property that can be associated with famous biological facts of genetic inheritance? Yes, such mutual property exists and it is connected with the tetra-reproduction by analogy with meiosis again. This property is not-trivial one and it does not exist in the most variants of arbitrary dispositions of 32 black triplets and 32 white triplets in (8*8)-matrices.

We have mentioned already that each mosaic row of the considered genetic matrices (see Figure 4) corresponds to one of Rademacher functions ($r_n(t) = \text{sign}(\sin 2^n \pi t)$, n = 1, 2, 3,…), which consists of components "+1" and "-1" only (see also [Petoukhov, arXiv:0802.3366v3]). Taking this fact into account, let us represent the black-and-white mosaic of each from the mentioned six genomatrices as a binary mosaic of numbers "+1" and "-1" by means of replacing black (white) color of each matrix cell by an element "+1" ("-1"). In such "Rademacher representation", the genomatrices $P^{CTAG}_{123}{}^{(3)}$, $P^{CTAG}_{231}{}^{(3)}$, $P^{CTAG}_{213}{}^{(3)}$, $P^{CTAG}_{321}{}^{(3)}$, $P^{CTAG}_{312}{}^{(3)}$, $P^{CTAG}_{132}{}^{(3)}$ are reformed into the genomatrices $B_{123}$, $B_{231}$, $B_{312}$, $B_{132}$, $B_{213}$, $B_{321}$ (Figure 7). Unexpected mutual property of these six binary genomatrices on Figure 7 is the following one. The multiplication of each genomatrix with itself (the square of each genomatrix) gives a phenomenon of its tetra-reproduction: four duplicates of the genomatrix are appeared. Really the following formulas take place:

$$(B_{123})^2 = 4*B_{123}; \quad (B_{231})^2 = 4*B_{231}; \quad (B_{312})^2 = 4*B_{312}$$

$$(B_{132})^2 = 4*B_{132}; \quad (B_{213})^2 = 4*B_{213}; \quad (B_{321})^2 = 4*B_{321} \quad (1)$$

This fact is interesting because the genetic code is destined by the nature for reproduction of biological structures, and matrices of the genetic code in their binary representation possess the non-trivial algebraic property of their own tetra-self-reproduction.

The set of these six binary genomatrices has many other interesting properties (for instance, $B_{123}*B_{321}+B_{123}*B_{132} = B_{123}{}^2$, etc.), which generate heuristic associations with genetic phenomena and which can be utilized to model the meiosis process of tetra-reproduction of gametal cells with a specific behavior of chromosomes. But these properties do not considered in this article.

It can be mentioned additionally that one can consider those "complementary" variants of the genomatrices $P^{CTAG}_{123}{}^{(3)}$ (or $P^{(3)}$ in Figure 1), $P^{CTAG}_{231}{}^{(3)}$, $P^{CTAG}_{213}{}^{(3)}$, $P^{CTAG}_{321}{}^{(3)}$, $P^{CTAG}_{312}{}^{(3)}$, $P^{CTAG}_{132}{}^{(3)}$, which are received by replacement of each triplet by its complementary triplet (the example of the complementary triplets is CAG and GTC). In each case the "complementary" matrix is identical to 180-degree turn of the initial matrix. The "complementary" genomatrices in similar binary presentations have the same properties of their tetra-reproduction [Petoukhov, 2008,b].

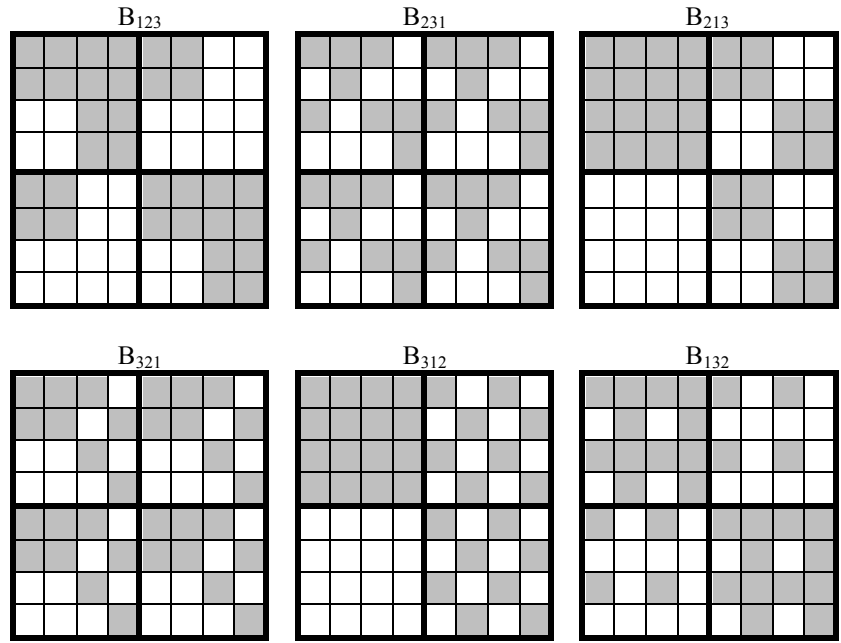

*Figure 7. Binary numeric genomatrices, in which each black cell means the element "+1"; each white cell means the element "-1".*

## 7 Projector operators in matrix presentations of the genetic code

The "Rademacher representation" of the genetic matrices on Figure 7 reveals also an unexpected connection of the genetic code structures with so called "projector operators", which are well-known in quantum mechanics, automatic control systems, etc. By definition, a linear operator P in a linear space is named a projector operator if it satisfies the following condition:

$$P^2 = P \qquad (2)$$

Let us consider matrices $Y_{123} = 4^{-1}*B_{123}$, $Y_{231} = 4^{-1}*B_{231}$, $Y_{312} = 4^{-1}*B_{312}$, $Y_{321} = 4^{-1}*B_{321}$, $Y_{213} = 4^{-1}*B_{213}$, $Y_{132} = 4^{-1}*B_{132}$. One can check that each of these matrices $Y_k$ is a projector operator because it satisfies the condition (2). We will name projector operators, which are based on genetic matrices, as "genoprojector operators". Some of these genoprojectors are commutative, other ones are non-commutative. Some examples of commutative genoprojectors are the following: $Y_{312}*Y_{213}=Y_{213}*Y_{312}$, $Y_{123}*Y_{321}=Y_{321}*Y_{123}$, $Y_{231}*Y_{132}=Y_{132}*Y_{231}$. Of course, each of new matrices $Y_{123}*Y_{321}$, $Y_{231}*Y_{132}$, $Y_{312}*Y_{213}$ is a projector also.

By definition, two projectors $P_1$ and $P_2$ are named orthogonal if $P_1*P_2=0$. One can check that the three genoprojectors $Y_{312}*Y_{213}$, $Y_{123}*Y_{321}$, $Y_{231}*Y_{132}$ are orthogonal each to other.

Some other variants of genetic (8*8)-matrices of triplets in the similar „Rademacher representation" with the same general factor $4^{-1}$ are genoprojector operators also: $[G\ A;\ T\ C]^{(3)}$, $[G\ T;\ A\ C]^{(3)}$, $[C\ A;\ T\ G]^{(3)}$. These matrices contain C and G along their diagonals. All possible variants of permutations of positions of triplets (1-2-3, 2-3-1, etc) in these matrices lead to new genoprojector operators. But the genomatrices, which contain the letters C and G not along their diagonals, do not lead to projector operators: $[C\ T;\ G\ A]^{(3)}$, $[C\ G;\ A\ T]^{(3)}$, etc. The (8*8)-matrices of genoprojectors are connected with genetic 8-dimensional Yin-Yang algebras (or bipolar algebras) which were presented in our works [Petoukhov, arXiv:0803.3330 and arXiv:0805.4692; Petoukhov, 2008b; Petoukhov, He, 2009].

A similar situation is true for mosaic (4*4)-matrices of duplets where black (white) cells contain two first letters of the subsets of black (white) triplets (see Figure 8): $[C\ A;\ T\ G]^{(2)}$, $[G\ A;\ T\ C]^{(2)}$, $[G\ T;\ A\ C]^{(2)}$, $[C\ T;\ A\ G]^{(2)}$. The similar numeric presentation of these genomatrices (where black cells and white cells contain elements „+1" and „-1" correspondingly, with the general factor $2^{-1}$ in this case) lead to new genoprojectors. The permutation of positions in all duplets (2-1 instead 1-2) lead to new genoprogectors again (these genoprojectors are non-commutative). But this is true only for genomatrices which contain C and G along their diagonals. Other genomatrices, which contain C and G not along their diagonals, do not lead to projector operators.

It is known that theory of projector operators has a set of interesting results and applications (see for example [Halmos, 1974; Messiah, 1999]). For example every family of commutative projectors generates algorithmically a Boolean algebra of projectors. Taking this into account, questions about a set of genoprojectors and corresponding Boolean algebras of genoprojectors should be studied systematically in theoretical and applied aspects including their biological meaning. The phenomenological fact of the connection of genetic code structures with special classes of projective operators allows developing new methods in bioinformatics and in technological fields of genetic algorithms on the base of the theory of projectors. Specifically we study applications of the theory of projectors for analysis of genetic sequences.

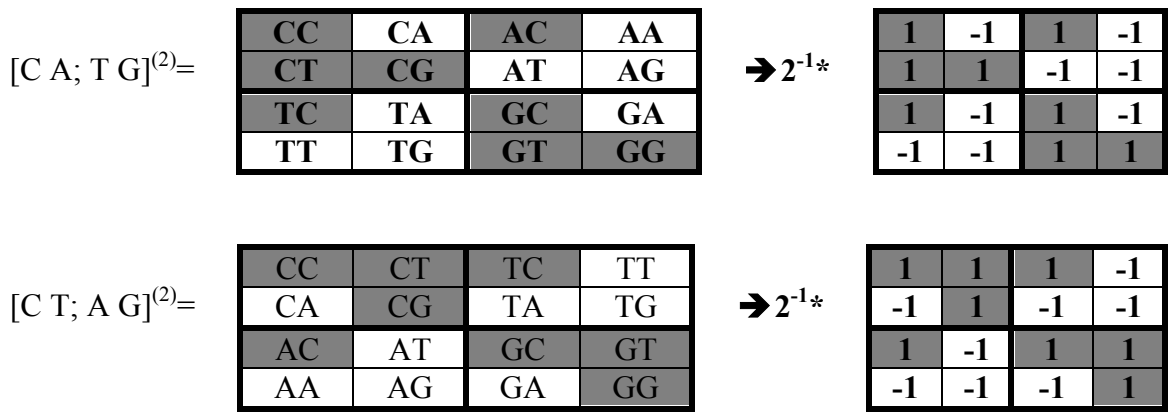

*Figure 8. Some examples of matrix projector operators (the numeric matrices on the right side) that correspond to mosaic matrices of duplets.*

**8 Unitary symmetries in matrix presentations of the genetic code**

Now we will pay our attention again to the fact that the genetic code is the bearer of the three pairs of binary-oppositional attributes which are showed on Figure 2. One can see that for a determination of any of genetic letters inside the genetic alphabet it is enough to indicate two from these three kinds of the mentioned binary-oppositional attributes. For example let us take the first two kinds of the binary-oppositional attributes from Figure 2. Let us mark each genetic letter with the amino-mutating property by the symbol "+1" (such genetic letter is amino from the viewpoint of its radical content), and each genetic letter without this property by the symbol "-1" (such genetic letter is keto from the viewpoint of its radical content). Let us mark else each pyrimidine by the symbol "+i" and each purine by the symbol "-i". Figure 9 shows what kind of pairs of these symbols (and of these attributes) determines each genetic letter.

|    | +i | -i |
|----|----|----|
| +1 | C  | A  |
| -1 | T  | G  |

Figure 9. The individual determination of each genetic letter by means of the individual pair of the attributes (explanation in text)

So, each genetic letter can be considered as a 2-parametric object inside the genetic alphabet. In view of this we put forward the following working hypothesis: one can reveal useful knowledge in mathematical bioinformatics (or in mathematical modeling of genetic systems) if each of the genetic letters is interpreted in the field of matrix genetics as a relevant 2-dimensional complex number:

$$C = 1+i, \; A = 1-i, \; T = -1+i, \; G = -1-i. \qquad (3)$$

The author has received some interesting results in matrix genetics on the base of this hypothesis (or of this numeric approach). One of them is a natural presentation of some families of genetic matrices as families of unitary matrices. This result seems to be interesting from the viewpoint of importance of unitary symmetries in theoretical physics [Lichtenberg, 1978; Louck, 2008; Rumer, Fet, 1970].

How many (2x2)-matrices can be constructed under the condition that each row and each column of each matrix contains only those genetic letters that belong to one of the mentioned binary-oppositional attributes (amino-keto and pyrimidine-purine)? In other words, each row and each column of such matrices should consist of the pairs C-A, C-T, G-A, G-T only. This condition is identical with the condition that such matrices should contain complementary pairs of the letters C-G and A-T along their diagonals (this condition exists already in the previous paragraph 7 about genetic projectors). One can check that only 8 symbolic matrices $P^{[CATG]}$, $P^{[CTAG]}$, $P^{[GATC]}$, $P^{[GTAC]}$, $P^{[ACGT]}$, $P^{[TCGA]}$, $P^{[AGCT]}$, $P^{[TGCA]}$ on Figure 10 satisfy this condition.

$$P^{[CATG]} = \begin{vmatrix} C & A \\ T & G \end{vmatrix} \rightarrow W_1 = (1+i)*8^{-0.5}* \begin{vmatrix} (1+i) & (1-i) \\ (-1+i) & (-1-i) \end{vmatrix}, \det(W_1) = 1$$

$$P^{[CTAG]} = \begin{vmatrix} C & T \\ A & G \end{vmatrix} \rightarrow W_2 = (1+i)*8^{-0.5}* \begin{vmatrix} (1+i) & (-1+i) \\ (1-i) & (-1-i) \end{vmatrix}, \det(W_2) = 1$$

$$P^{[GATC]} = \begin{vmatrix} G & A \\ T & C \end{vmatrix} \rightarrow W_3 = (1+i)*8^{-0.5}* \begin{vmatrix} (-1-i) & (1-i) \\ (-1+i) & (1+i) \end{vmatrix}, \det(W_3) = 1$$

$$P^{[GTAC]} = \begin{vmatrix} G & T \\ A & C \end{vmatrix} \rightarrow W_4 = (1+i)*8^{-0.5}* \begin{vmatrix} (-1-i) & (-1+i) \\ (1-i) & (1+i) \end{vmatrix}, \det(W_4) = 1$$

$$P^{[ACGT]} = \begin{vmatrix} A & C \\ G & T \end{vmatrix} \rightarrow W_5 = (1+i)*8^{-0.5}* \begin{vmatrix} (1-i) & (1+i) \\ (-1-i) & (-1+i) \end{vmatrix}, \det(W_5) = -1$$

$$P^{[TCGA]} = \begin{vmatrix} T & C \\ G & A \end{vmatrix} \rightarrow W_6 = (1+i)*8^{-0.5}* \begin{vmatrix} (-1+i) & (1+i) \\ (-1-i) & (1-i) \end{vmatrix}, \det(W_6) = -1$$

$$P^{[AGCT]} = \begin{vmatrix} A & G \\ C & T \end{vmatrix} \rightarrow W_7 = (1+i)*8^{-0.5}* \begin{vmatrix} (1-i) & (-1-i) \\ (1+i) & (-1+i) \end{vmatrix}, \det(W_7) = -1$$

$$P^{[TGCA]} = \begin{vmatrix} T & G \\ C & A \end{vmatrix} \rightarrow W_8 = (1+i)*8^{-0.5}* \begin{vmatrix} (-1+i) & (-1-i) \\ (1+i) & (1-i) \end{vmatrix}, \det(W_8) = -1$$

Figure 10. The algorithmic presentation of the eight symbolic genomatrices $P^{[CATG]}$, $P^{[CTAG]}$, $P^{[GATC]}$, $P^{[GTAC]}$, $P^{[ACGT]}$, $P^{[TCGA]}$, $P^{[AGCT]}$, $P^{[TGCA]}$ in the form of numeric matrices $W_k$ (k=1, 2, …., 8), which are unitary matrices.

Let us replace symbols C, A, T, G in these 8 matrices by means of their complex presentations from expressions (3). Additionally each numeric matrix is normalized by means of its multiplication with the factor $(1+i)*8^{-0.5}$ to provide the value $\pm 1$ for the matrix determinant. Figure 10 shows the result of this action in a form of numeric matrices.

By definition a unitary matrix is a complex (n*n)-matrix W satisfying the condition

$$W*\hat{W} = \hat{W}*W = E_n \qquad (4)$$

Where $E_n$ is the identity matrix in n dimensions and $\hat{W}$ is the conjugate transpose of W (also called the Hermitian adjoint). One can check that all eight matrices $W_k$ (k = 1, 2, …., 8) on Figure 10 satisfy the condition (4). So they are unitary matrices.

Determinants of the first four matrices $W_1$, $W_2$, $W_3$, $W_4$ are equal to "+1". Determinants of the last four matrices $W_5$, $W_6$, $W_7$, $W_8$ are equal to "-1". Each of the first four matrices $W_1$, $W_2$, $W_3$, $W_4$ is the basis of the cyclic group of unitary transformations $(W_m)^s$ where m = 1, 2, 3, 4 and "s" is integer number. The period of these cyclic groups is equal to 4: $(W_m)^s = (W_m)^{s+4}$. Each of the last four matrices $W_5$, $W_6$, $W_7$, $W_7$ is the basis of the cyclic group of unitary transformations $(W_n)^s$ where n = 5, 6, 7, 8 and "s" is integer number. The period of these cyclic groups is equal to 2: $(W_n)^s = (W_n)^{s+2}$.

It is known that Kronecker product of unitary matrices is creating a new unitary matrix. Different Kronecker products of all these 8 unitary matrices $W_k$ (k = 1, 2, …., 8) on Figure 10 create new unitary matrices. For example each matrix $W_k^{(3)}$ is the unitary (8*8)-matrix. Each family of $(W_k^{(3)})^s$, where s is integer number, gives a cyclic group of unitary transformations also.

Some variants of permutations of genetic elements in genetic matrices lead to new unitary genetic matrices by analogical algorithm. These new unitary genomatrices are generating their new cyclic groups of unitary transformations.

It should be noted that the idea of presentation of the genetic letters in a form of complex numbers is not the new idea. For example, a presentation of genetic letters by means of complex numbers was made in the work (Cristea, 2005). But our approach has one very important aspect: we use such presentation for matrix analysis of ensemble of genetic multiplets. In the result the connection of the genetic code with unitary symmetries was discovered. Such matrix approach was not done by other authors as we know.

In the beginning of this section we have taken the pair of binary-oppositional attributes "amino-keto" and "pyrimidine-purine" to construct Figure 9 with the expression (3). But one can take another pair of the binary-oppositional attributes from Figure 2, for example the attribute "amino-keto" and the attribute "two hydrogen bonds – three hydrogen bonds". By analogy in this case one can make another presentation of the genetic letters in a form of complex numbers to consider another genetic (2*2)-matrices [C  A; G  T], [A  C; T  G], etc. where the pyrimidine C-T belong to one matrix diagonal and the purine A-G belong to another matrix diagonal. It is obvious that analogical unitary matrices are generated in this case. We do not consider such cases separately because they are identical to the previous case in general from the formal viewpoint.

### 9. About complementary palindromes, reproductions and the Rademacher genomatrices

One of important problems in molecular genetics is the problem of "complementary palindromes" in nucleotide sequences. In linguistics the notion "palindrome" means a row that is read identically in both directions (the words *civic*, *radar*, *level* are examples of linguistic palindromes). In genetics the notion "complementary palindrome" is important. By definition, a complementary palindrome is a row of DNA or RNA that becomes a simple palindrome if each symbol in one of halves of this row is replaced by its complementary symbol (in DNA the complementary pairs are A-T and C-G, and in RNA complementary pairs are A-U and C-G). For example CTCGCGAG is a complementary palindrome in DNA. Many complementary palindromes exist in DNA sequences. For example the fourth chromosome of the genome Arabidopsis th. contains the complementary palindrome TGTCGATCGACA which is repeated 194 times there [Gusev et al, 2009]. The problem of complementary palindromes in genetics is considered in many works (see for example [Gusev et al., 2009; Gusfield, 1999]).

This phenomenological problem has some unexpected relations with mathematical properties of the genomatrices [C T; A G]$^{(2)}$ and [C T; A G]$^{(3)}$ (Figure 1). On the base of these relations, mathematical models can be developed for a description of realizations of $2^n$-dimensional vectors of the complementary-palindromic type together with a description of $2^n$-reproductions of such $2^n$-dimensional vectors. Let us consider these relations more attentively.

Figure 11 shows the genomatrix [C T; A G]$^{(3)}$ of 64 triplets where all the triplets with strong roots CC, CT, CG, AC, TC, GC, GT, GG are marked by black color. Figure 11 shows also the genomatrix [C A; T G]$^{(2)}$ of 16 duplets where all of these "strong" duplets CC, CT, CG, AC, TC, GC, GT, GG are marked by black color as well. From the viewpoint of the black-and-white mosaics, each row of these genomatrices corresponds to one of Rademacher functions which contains only elements "+1" and "-1". Figures 11 shows the Rademacher representations $R_4$ and $R_8$ of these genomatrices also.

|  | CC | CT | TC | TT |
|---|---|---|---|---|
| [C T; A G]$^{(2)}$= | CA | CG | TA | TG |
|  | AC | AT | GC | GT |
|  | AA | AG | GA | GG |

|  | 1 | 1 | 1 | -1 |
|---|---|---|---|---|
| $R_4$ = | -1 | 1 | -1 | -1 |
|  | 1 | -1 | 1 | 1 |
|  | -1 | -1 | -1 | 1 |

| CCC | CCT | CTC | CTT | TCC | TCT | TTC | TTT |
|---|---|---|---|---|---|---|---|
| CCA | CCG | CTA | CTG | TCA | TCG | TTA | TTG |
| CAC | CAT | CGC | CGT | TAC | TAT | TGC | TGT |
| CAA | CAG | CGA | CGG | TAA | TAG | TGA | TGG |
| ACC | ACT | ATC | ATT | GCC | GCT | GTC | GTT |
| ACA | ACG | ATA | ATG | GCA | GCG | GTA | GTG |
| AAC | AAT | AGC | AGT | GAC | GAT | GGC | GGT |
| AAA | AAG | AGA | AGG | GAA | GAG | GGA | GGG |

$R_8$=

| 1 | 1 | 1 | 1 | 1 | 1 | -1 | -1 |
|---|---|---|---|---|---|---|---|
| 1 | 1 | 1 | 1 | 1 | 1 | -1 | -1 |
| -1 | -1 | 1 | 1 | -1 | -1 | -1 | -1 |
| -1 | -1 | 1 | 1 | -1 | -1 | -1 | -1 |
| 1 | 1 | -1 | -1 | 1 | 1 | 1 | 1 |
| 1 | 1 | -1 | -1 | 1 | 1 | 1 | 1 |
| -1 | -1 | -1 | -1 | -1 | -1 | 1 | 1 |
| -1 | -1 | -1 | -1 | -1 | -1 | 1 | 1 |

Figure 11 The genomatrices [C T; A G]$^{(2)}$ (in the upper row) and [C T; A G]$^{(3)}$ and their Rademacher representations $R_4$ and $R_8$ are shown. Black cells contain strong duplets CC, CT, CG, AC, TC, GC, GT, GG and triplets with these strong roots. Each column in $R_4$ and $R_8$ corresponds to one of Rademacher functions.

These Rademacher genomatrices $R_4$ and $R_8$ possess interesting properties. Firstly, their actions on the arbitrary 4-dimensional and 8-dimensional vectors $V_4$ = [$a_0$ $a_1$ $a_2$ $a_3$] and $V_8$ = [$a_0$ $a_1$ $a_2$ $a_3$ $a_4$ $a_5$ $a_6$ $a_7$] correspondingly are leading to new vectors in a form of complementary palindromes (Figure 12). Each of these new vectors becomes a simple palindrome by means of inversion of signs of all components in one of two its halves. In the case of 8-dimensional vectors on Figure 12, new palindromic vectors possess an additional peculiarity: each pair of adjacent members in such 8-dimensional palindromic vectors contains identical components.

| 4-dimensional vectors | Numeric example | $R_4$*[3, -8, 5, 7]$^T$ = [-7, -23, 23, 7]$^T$ |
|---|---|---|
|  | General case | $R_4$*[$a_0$, $a_1$, $a_2$, $a_3$]$^T$ = [$a_0$+$a_1$+$a_2$-$a_3$, -$a_0$+$a_1$-$a_2$-$a_3$, $a_0$-$a_1$+$a_2$+$a_3$, -$a_0$-$a_1$-$a_2$+$a_3$]$^T$ |
| 8-dimensional vectors | Numeric example | $R_8$*[3 -8 5 7 6 9 1 -4]$^T$ = = [25 25 55 -5 -5 -25 -25]$^T$ |
|  | General case | $R_8$*[$a_0$ $a_1$ $a_2$ $a_3$ $a_4$ $a_5$ $a_6$ $a_7$]$^T$ = [$a_0$+$a_1$+$a_2$+$a_3$+$a_4$+$a_5$-$a_6$-$a_7$, $a_0$+$a_1$+$a_2$+$a_3$+$a_4$+$a_5$-$a_6$-$a_7$, -$a_0$-$a_1$+$a_2$+$a_3$-$a_4$-$a_5$-$a_6$-$a_7$, -$a_0$-$a_1$+$a_2$+$a_3$-$a_4$-$a_5$-$a_6$-$a_7$, $a_0$+$a_1$-$a_2$-$a_3$+$a_4$+$a_5$+$a_6$+$a_7$, $a_0$+$a_1$-$a_2$-$a_3$+$a_4$+$a_5$+$a_6$+$a_7$, -$a_0$-$a_1$-$a_2$-$a_3$-$a_4$-$a_5$+$a_6$+$a_7$, -$a_0$-$a_1$-$a_2$-$a_3$-$a_4$-$a_5$+$a_6$+$a_7$]$^T$ |

Figure12. Transformations of arbitrary 4-dimensional and 8-dimensional vectors into vectors of the complementary-palindrome type. Complementary pairs of components in both halves of the vectors of the complementary-palindrome type are marked by identical colors. The symbol T over vectors means transpose of these vectors.

The second heuristic property of such Rademacher genomatrices $R_4$ and $R_8$ is the following. A repeating action of the genomatrix $R_4$ on any complementary-palindrome vector [$a_0$, $a_1$, -$a_1$, -$a_0$]$^T$ generates a dichotomous reproduction of this vector:

$$R_4*[a_0, a_1, -a_1, -a_0]^T = 2*[a_0, a_1, -a_1, -a_0] \qquad (5)$$

It can be used as a model of inherited dichotomous reproductions of biological cells in a course of mitosis when genetic materials are reproduced in dichotomous manner. One can mention else that each column of the Rademacher genomatrix $R_4$ is a complementary-palindrome vector itself. By this reason the exponentiation of this genomatrix generates its dichotomic reproduction also: $R_4^2 = \mathbf{2}*R_4$.

A similar situation exists for the Rademacher genomatrix $R_8$:

$$R_8*(R_8*[a_0\ a_1\ a_2\ a_3\ a_4\ a_5\ a_6\ a_7]^T) = \mathbf{4}*R_8*[a_0\ a_1\ a_2\ a_3\ a_4\ a_5\ a_6\ a_7]^T \qquad (6)$$

It can be used as a model of inherited tetra-reproductions of germ cells in a course of meiosis when genetic materials are reproduced in tetra-reproduction manner (four germ cells arise from one germ cell in a course of meiosis). Each column of the Rademacher genomatrix $R_8$ is a complementary-palindrome vector itself. By this reason the exponentiation of this genomatrix generates its tetra-reproduction also: $R_8^2 = \mathbf{4}*R_8$.

The author's works [Petoukhov, 2011, 2012a] describes other interesting properties of these Rademacher genomatrices $R_4$ and $R_8$ which are close connected with special types of 4-dimensional and 8-dimensional hypercomplex numbers well-known in physics: split-quaternions and bi-split-quaternions by J.Cockle.

## 10 Transformations of the genomatrix [C T; A G]$^{(3)}$ by removing separate positions in the triplets

The mosaic genomatrix [C T; A G]$^{(3)}$ (Figure 11) possesses interesting properties relative to removing separate positions in its triplets (this method provides a positional convolution of each triplet into a corresponding duplet). This operation leads to new mosaic genomatrices (Figures 13-15) where the strong roots (the duplets CC, CT, CG, AC, TC, GC, GT, GG) are again marked by black color and the weak roots (CA, AA, AT, AG, TA, TT, TG, GA) are marked by white color. Removing the first positions of triplets in [C T; A G]$^{(3)}$ leads to the (8*8)-genomatrix of duplets (Figure 13, left) where each column corresponds to one of Rademacher functions relative to its black-and-white mosaic. Removing the second positions of triplets in [C T; A G]$^{(3)}$ leads to the (8*8)-genomatrix of duplets (Figure 14, left) where each column corresponds again to one of Rademacher functions relative to its black-and-white mosaic. Removing the third positions of triplets in [C T; A G]$^{(3)}$ leads to the (8*8)-genomatrix of duplets (Figure 15, left) where each column corresponds again to one of Rademacher functions relative to its black-and-white mosaic.

The Rademacher representations R1, R2 and R3 (Figure 13-15, right sides) of these new genomatrices and the Rademacher representation $R_8$ of the initial genomatrix [C T; A G]$^{(3)}$ possess some similar properties. For example, the exponentiation of these genomatrices $R_8$, R1, R2 and R3 generates their tetra-reproduction: $R_8^2=\mathbf{4}*R_8$, $R1^2=\mathbf{4}*R1$, $R2^2=\mathbf{4}*R2$ and $R3^2=\mathbf{4}*R3$. The dyadic-shift decomposition of each of these Rademacher genomatrices R1, R2 and R3 generates an individual set of 8 sparse matrices; each of these sets is closed relative to multiplication and its multiplication table is identical to the multiplication table of 8-dimensional hypercomplex numbers which are well-known in mathematics and physics under the name bi-split-quaternions by J.Cockle (see details in our works [Petoukhov, 2011, 2012a]).

| CC | CT | TC | TT | CC | CT | TC | TT |
|----|----|----|----|----|----|----|----|
| CA | CG | TA | TG | CA | CG | TA | TG |
| AC | AT | GC | GT | AC | AT | GC | GT |
| AA | AG | GA | GG | AA | AG | GA | GG |
| CC | CT | TC | TT | CC | CT | TC | TT |
| CA | CG | TA | TG | CA | CG | TA | TG |
| AC | AT | GC | GT | AC | AT | GC | GT |
| AA | AG | GA | GG | AA | AG | GA | GG |

| 1 | 1 | 1 | -1 | 1 | 1 | 1 | -1 |
|---|---|---|----|---|---|---|----|
| -1 | 1 | -1 | -1 | -1 | 1 | -1 | -1 |
| 1 | -1 | 1 | 1 | 1 | -1 | 1 | 1 |
| -1 | -1 | -1 | 1 | -1 | -1 | -1 | 1 |
| 1 | 1 | 1 | -1 | 1 | 1 | 1 | -1 |
| -1 | 1 | -1 | -1 | -1 | 1 | -1 | -1 |
| 1 | -1 | 1 | 1 | 1 | -1 | 1 | 1 |
| -1 | -1 | -1 | 1 | -1 | -1 | -1 | 1 |

Figure 13. Left: the transformation of the genomatrix [C T; A G]$^{(3)}$ (Figure 11) by means of removing the first position in each triplets. Right: the Rademacher form R1 of this new genomatrix (black and white cells contain elements +1 and -1 correspondingly). Explanations in the text.

| CC | CT | CC | CT | TC | TT | TC | TT |
|----|----|----|----|----|----|----|----|
| CA | CG | CA | CG | TA | TG | TA | TG |
| CC | CT | CC | CT | TC | TT | TC | TT |
| CA | CG | CA | CG | TA | TG | TA | TG |
| AC | AT | AC | AT | GC | GT | GC | GT |
| AA | AG | AA | AG | GA | GG | GA | GG |
| AC | AT | AC | AT | GC | GT | GC | GT |
| AA | AG | AA | AG | GA | GG | GA | GG |

| 1 | 1 | 1 | 1 | 1 | -1 | 1 | -1 |
|---|---|---|---|---|----|---|----|
| -1 | 1 | -1 | 1 | -1 | -1 | -1 | -1 |
| 1 | 1 | 1 | 1 | 1 | -1 | 1 | -1 |
| -1 | 1 | -1 | 1 | -1 | -1 | -1 | -1 |
| 1 | -1 | 1 | -1 | 1 | 1 | 1 | 1 |
| -1 | -1 | -1 | -1 | -1 | 1 | -1 | 1 |
| 1 | -1 | 1 | -1 | 1 | 1 | 1 | 1 |
| -1 | -1 | -1 | -1 | -1 | 1 | -1 | 1 |

Figure 14. Left: the transformation of the genomatrix [C T; A G]$^{(3)}$ (Figure 11) by means of removing the second position in each triplets. Right: the Rademacher form R2 of this new genomatrix (black and white cells contain elements +1 and -1 correspondingly). Explanations in the text.

| CC | CC | CT | CT | TC | TC | TT | TT |
|----|----|----|----|----|----|----|----|
| CC | CC | CT | CT | TC | TC | TT | TT |
| CA | CA | CG | CG | TA | TA | TG | TG |
| CA | CA | CG | CG | TA | TA | TG | TG |
| AC | AC | AT | AT | GC | GC | GT | GT |
| AC | AC | AT | AT | GC | GC | GT | GT |
| AA | AA | AG | AG | GA | GA | GG | GG |
| AA | AA | AG | AG | GA | GA | GG | GG |

| 1 | 1 | 1 | 1 | 1 | 1 | -1 | -1 |
|---|---|---|---|---|---|----|----|
| 1 | 1 | 1 | 1 | 1 | 1 | -1 | -1 |
| -1 | -1 | 1 | 1 | -1 | -1 | -1 | -1 |
| -1 | -1 | 1 | 1 | -1 | -1 | -1 | -1 |
| 1 | 1 | -1 | -1 | 1 | 1 | 1 | 1 |
| 1 | 1 | -1 | -1 | 1 | 1 | 1 | 1 |
| -1 | -1 | -1 | -1 | -1 | -1 | 1 | 1 |
| -1 | -1 | -1 | -1 | -1 | -1 | 1 | 1 |

Figure 15. Left: the transformation of the genomatrix [C T; A G]$^{(3)}$ (Figure 11) by means of removing the third position in each triplets. Right: the Rademacher form R3 of this new genomatrix (black and white cells contain elements +1 and -1 correspondingly). Explanations in the text.

The stability of algebraic properties of these genomatrices adds materials to the author's idea about importance of "positional languages" for molecular genetics or, in other words, to the idea of positioning of the binary languages in genetic sequences (we are talking about reading each of the three positions inside triplets based on one of the three pairs of binary sub-alphabets from Figure 2) [Petoukhov S.V., 2003, p. 14; 2008; 2012a; 2012b; Petoukhov, He, 2010].

This method of a positional convolution of triplets and other n-plets (or oligonucleotides) will be used in the next section to construct fractal genetic nets (FGN) as a new tool to reveal hidden rules of long nucleotide sequences.

# 11 Fractal genetic nets and the Symmetry Principles of long nucleotide sequences

On the base of his results in the field of matrix genetics, the author proposes a new notion "fractal genetic nets" (FGN) which is a useful tool to study long genetic sequences, first of all, to study symmetrical properties of long nucleotide sequences. In general case each variant of FGN is constructed by means of the author's "method of a positional convolution of long genetic sequences" to get a bunch of long sequences, each of which, respectively, shorter than the original sequence. In the particular case, this method lies in the positional convolution of long sequences of triplets through the removal or retention of individual positions (items) in each triplet by an analogy to the steps in the previous section (Figures 13-15).

In literature sources, long genetic sequences are termed those that contain no less that 50.000 nucleotides (see for example [Yamagishi, Herai, 2011]). In results of our preliminary researches of long nucleotide sequences of organisms of various taxonomic types, we reveal evidences of the author's hypothesis: hidden regularities of long genetic sequences are connected with fractal genetic nets (FGN); studying of long genetic sequences by means of using FGN allows discovering new hidden rules of living nature.

Let us explain a construction of FGN of various types on an example of FGN for sequences of triplets (Figure 16). Three positions in each triplet are numerated by numbers 0, 1 and 2 correspondingly. At the first level of a convolution, an initial long sequence $S_0$ of triplets is transformed by means of a positional convolution into three new sequences of nucleotides $S_{1/0}$, $S_{1/1}$, $S_{1/2}$, each of which is shorter in 3 times in comparison with the initial sequence (numerator of the index in this notation of sequences shows the level of the convolution, and the denominator - the position of the triplets, which is used for the convolution): the sequence $S_{1/0}$ includes one by one all the nucleotides that are in the initial position "0" of triplets of the original sequence $S_0$; the sequence $S_{1/1}$ includes one by one all the nucleotides that are in the middle position "1" of triplets of the original sequence $S_0$; the sequence $S_{1/2}$ includes one by one all the nucleotides that are in the last position "2" of triplets of the original sequence $S_0$. At the final stage of the first level of the positional convolution, each of the sequences of nucleotides $S_{1/0}$, $S_{1/1}$, $S_{1/2}$ is represented as a sequence of triplets where three positions inside each triplets are numerated again by numbers 0, 1 and 2. To construct the second level of the convolution, each of the sequences $S_{1/0}$, $S_{1/1}$, $S_{1/2}$ is transformed by means of the same positional convolution in three new sequences: $S_{1/0}$ is convolved in $S_{2/00}$, $S_{2/01}$, $S_{2/02}$; $S_{1/1}$ – in $S_{2/10}$, $S_{2/11}$, $S_{2/12}$; $S_{1/2}$ – in $S_{2/20}$, $S_{2/21}$, $S_{2/22}$. The third level and subsequent levels of the convolution are constructed by analogy to form a multi-level net of sequences of nucleotides, which is termed "the fractal genetic net for the triplet convolution" or briefly "FGN-3" (Figure 16).

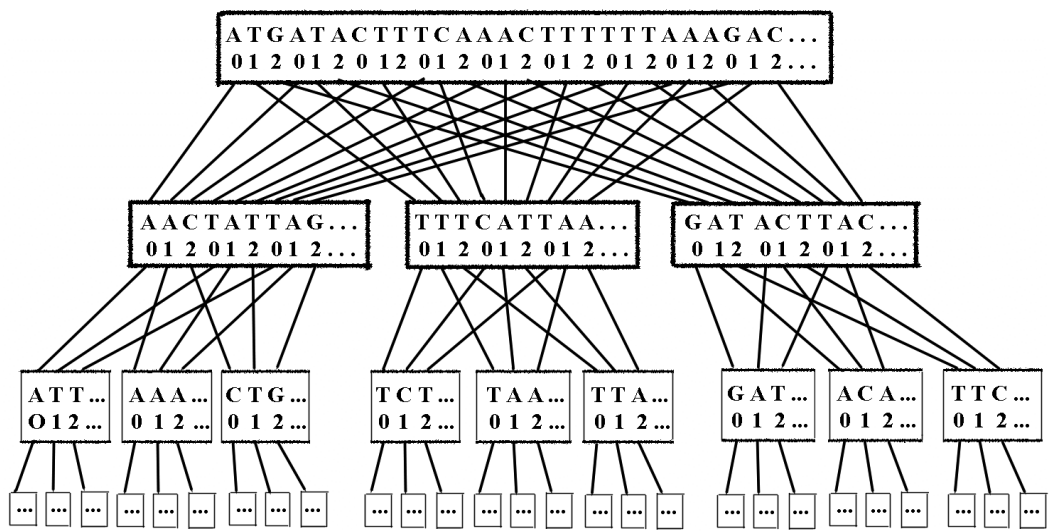

*Figure 16. The scheme of the fractal genetic net (FGN-3) for a sequence of triplets*

This FGN possesses a fractal-like character if the numeration of positions is only taken into account: each of long sequences of this FGN can be taken as an initial sequence to form a similar genetic net on its basis (Figure 17). In general case, the FGN can be built not only for triplets, but also for other n-plets (n = 2, 4, 5, ...) or oligonucleotides by means of a repeated positional convolution of each of sequences from the previous level into "n" sequences of the next level of the convolution. In this way one can built FGN-2, FGN-4, FGN-5, etc. for n=2, 3, 4, 5,... correspondingly. (Each of these FGN-2, FGN-3, FGN-4, FGN-5, etc. is a tree, but all of them form a net of separate trees; in a wide sense, FGN is the complete set of such separate trees). But this article describes some our results only for the FGN-3.

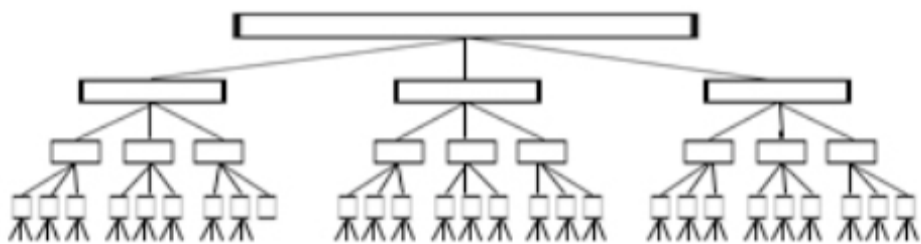

Figure 17. The fractal scheme of the triple branching in the case of the FGN-3

To test the author's hypothesis that structures of long nucleotide sequences of different organisms are connected with fractal genetic nets (first of all with FGN-3), we analyze an implementation of the known «Symmetry Principle» [Yamagishi, Herai, 2011, p.2] for long nucleotide sequences of different levels of a positional convolution in the fractal genetic net for the triplet convolution (FGN-3). Let us remind briefly about this Symmetry Principle which was studied or described in many publications [Bell, Forsdyke, 1999; Chargaff, 1971, 1975; Dong, Cuticchia, 2001; Forsdyke, 2002; Forsdyke, Bell, 2004; Kong, et al. 2009; Mitchell, Bridge, 2006; Prabhu, 1993; Sueoka, 1999; Yamagishi, Herai, 2011].

The Chargaff's first parity rule speaks that in any double-stranded DNA segment, the quantities (or frequencies) of adenine and thymine are equal, and so are the frequencies of cytosine and guanine [Chargaff, 1950]. This rule was used by Watson and Crick to support their famous DNA double-helix structure model [Watson & Crick, 1953]. Chargaff also perceived that the parity rule approximately holds in the single-stranded DNA segment. This last rule is known as Chargaff's second parity rule (CSPR), and it has been confirmed in several organisms [Mitchell & Bride, 2006]. Originally, CSPR is meant to be valid only to mononucleotide frequencies (that is quantities of monoplets) in the single-stranded DNA. *"But, it occurs that oligonucleotide frequencies follow a generalized Chargaff's second parity rule (GCSPR) where the frequency of an oligonucleotide is approximately equal to its complement reverse oligonucleotide frequency [Prahbu, 1993]. This is known in the literature as the Symmetry Principle"* [Yamagishi, Herai, 2011, p. 2]. The work [Prahbu, 1993] shows the implementation of the Symmetry Principle in long DNA-sequences for cases of complementary reverse n-plets with n = 2, 3, 4, 5 at least.

In our article we use two different notions of complementary oligonucleotides (or n-plets): 1) complementary oligonucleotides in a traditional sense (for example ACGTG and TGCAC are the pair of complementary oligonucleotides in a traditional sense); 2) complementary reverse oligonucleotides or briefly CR-oligonucleotides or reverse complements (for example ACGTG and CACGT are the pair of CR-oligonucleotides). The mentioned Symmetry Principle has been revealed for pairs of CR-oligonucleotides. Taking this into account we began testing the author's hypothesis by means of analyzing frequencies (or quantities) of all variants of pairs of CR-oligonucleotides in long DNA-sequences of different organisms at different levels of their FGN. We test frequencies of n-plets in the FGN-3 with n =

1, 2, 3, 4, 5 only because of our computer limitations, but we suppose that our described results for FGN-3 hold true also for n > 5. Initial nucleotide sequences for testing are taken from http://www.ncbi.nlm.nih.gov/. To test the author's hypothesis we use a special software written by V.I.Svirin on the basis of the computer language Python under the technical project by the author.

In the result of our preliminary studies we have revealed the following: 1) the Symmetry Principle for pairs of CR-oligonucleotides is realized in each of long nucleotide sequences at different levels of the convolution in FGN-3 (the length of oligonucleotides or n-plets under consideration is equal to n = 1, 2, 3, 4, 5 at least); 2) a series of new Symmetry Principles exists in those initial long nucleotide sequences where the famous Symmetry Principle for pairs of CR-oligonucleotides is performed; 3) each of these new Symmetry Principles is performed for n-plets in each of long nucleotide sequences at different levels of the convolution in FGN-3 (n = 1, 2, 3, 4, 5 at least).

Let us take for example the long nucleotide sequence of Mycoplasma crocodyli MP145 chromosome, complete genome (NCBI Reference Sequence: NC_014014.1 http://www.ncbi.nlm.nih.gov/nuccore/294155300). This sequence contains 934379 nucleotides. Figure 18 shows realisations of the known Symmetry Principle (we'll name it as the Symmetry Principle №1) in the 13 sequences of the first three levels of convolution in the FGN-3 of this genome. It displays the number of occurences of 32 triplets (AAA, AAC, AAG, AAT, ACA, ACC, ACG, ACT, AGA, AGC, AGG, ATA, ATC, ATG, CAA, CAC, CAG, CCA, CCC, CCG, CGA, CGC, CTA, CTC, GAA, GAC, GCA, GCC, GGA, GTA, TAA, TCA) and their 32 CR-triplets (TTT, GTT, CTT, ATT, TGT, GGT, CGT, AGT, TCT, GCT, CCT, TAT, GAT, CAT, TTG, GTG, CTG, TGG, GGG, CGG, TCG, GCG, TAG, GAG, TTC, GTC, TGC, GGC, TCC, TAC, TTA, TGA) in the long sequences $S_0$, $S_{1/0}$, $S_{1/1}$, $S_{1/2}$, $S_{2/00}$, $S_{2/01}$, $S_{2/02}$, $S_{2/10}$, $S_{2/11}$, $S_{2/12}$, $S_{2/20}$, $S_{2/21}$, $S_{2/22}$ at the first three levels of the FGN-3 (a limited volume of the article doesn't allow showing other levels of this FGN).

The straight line in each frame is of slope 1 (it is a bisector of the coordinate angle). Each dot in a frame represents one pair "triplet and CR-triplet"; its coordinate X shows number of occurences ( or the frequence) of the triplet, and its coordinate Y shows number the frequence of its CR-triplet on the same strand of the sequence. Each frame contains all 32 pairs «triplet and its CR-triplet». The dots agglutinate at the line of slope 1, demonstrating that amounts of occurrences (or frequences) of two members of each of 32 pairs «triplet and its CR-triplet» are aproximately equal in each of the sequences at each of the levels of convolution in the FGN-3. It means that the Symmetry Principle №1 is performed for each of these sequences.

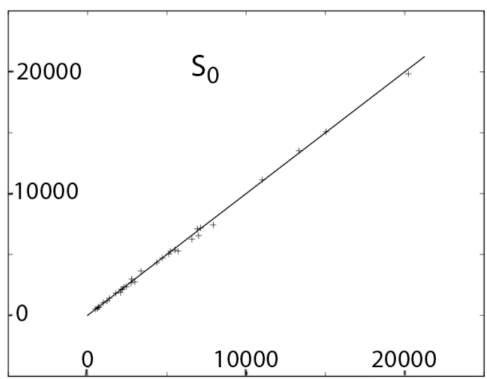

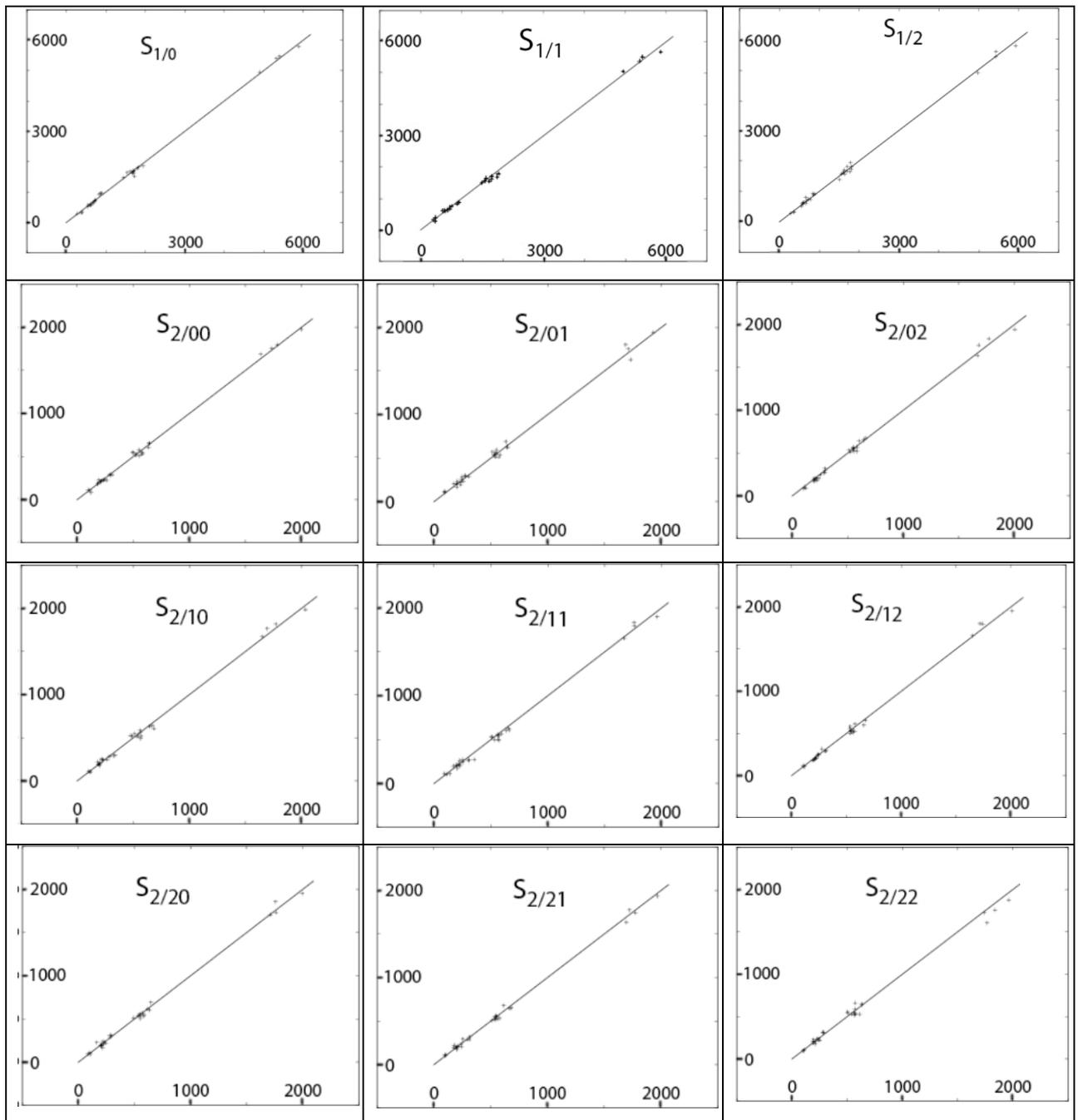

Figure 18. Realizations of the Symmetry Principle №1 in the long sequences $S_0$, $S_{1/0}$, $S_{1/1}$, $S_{1/2}$, $S_{2/00}$, $S_{2/01}$, $S_{2/02}$, $S_{2/10}$, $S_{2/11}$, $S_{2/12}$, $S_{2/20}$, $S_{2/21}$, $S_{2/22}$ at the first three levels of the FGN-3 for Mycoplasma crocodyli MP145 chromosome, complete genome (NCBI Reference Sequence: NC_014014.1 http://www.ncbi.nlm.nih.gov/nuccore/294155300). The initial sequence $S_0$ contains 934379 nucleotides.

On this basis the author notes existence of
**the generalized Symmetry Principle № 1**:
- in long nucleotide sequences at different levels of convolution in FGN-3, oligonucleotide frequencies follow a generalized Chargaff's second parity rule where the frequency of each oligonucleotide is approximately equal to its complement reverse oligonucleotide frequency.

Now let us describe new Symmetry Principles discovered by us for long nucleotide sequences in a connection with obtained results of matrix genetics.

**The Symmetry Principle № 2** (concerning to FGN): the frequency of each oligonucleotide is approximately the same in all the long nucleotide sequences of each of levels of FGN-3.

Figure 19 shows an example of frequencies of the triplet ACG in 40 long nucleotide sequences $S_0$, $S_{1/0}$, $S_{1/1}$, $S_{1/2}$, $S_{2/00}$, $S_{2/01}$, $S_{2/02}$, $S_{2/10}$, $S_{2/11}$, $S_{2/12}$, $S_{2/20}$, $S_{2/21}$, $S_{2/22}$, ….., $S_{3/221}$, $S_{3/222}$ at the first four levels of the FGN-3 of the same initial sequences as on Figure 18.

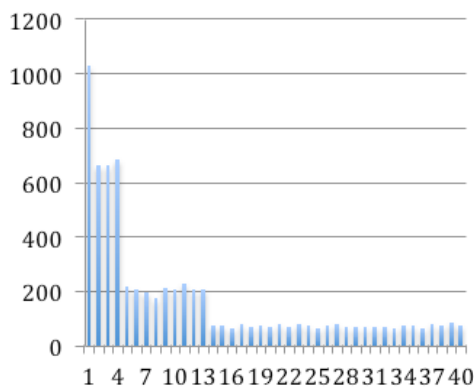

Figure 19. Frequencies of the triplet ACG in 40 long nucleotide sequences $S_0$, $S_{1/0}$, $S_{1/1}$, $S_{1/2}$, $S_{2/00}$, $S_{2/01}$, $S_{2/02}$, $S_{2/10}$, $S_{2/11}$, $S_{2/12}$, $S_{2/20}$, $S_{2/21}$, $S_{2/22}$, ….., $S_{3/221}$, $S_{3/222}$ at the first four levels of the FGN-3 of Mycoplasma crocodyli MP145 chromosome, complete genome (NCBI Reference Sequence: NC_014014.1 http://www.ncbi.nlm.nih.gov/nuccore/294155300). Coordinates X show the 40 sequences and coordinate Y show appropriate frequencies of the triplet ACG in them.

Figure 20 shows examples of frequencies of all 64 triplets in 13 long nucleotide sequences at the first three levels of FGN-3 of the same genome.

|     | $S_0$ | $S_{1/0}$ | $S_{1/1}$ | $S_{1/2}$ | $S_{2/00}$ | $S_{2/01}$ | $S_{2/02}$ | $S_{2/10}$ | $S_{2/11}$ | $S_{2/12}$ | $S_{2/20}$ | $S_{2/21}$ | $S_{2/22}$ |
| --- | --- | --- | --- | --- | --- | --- | --- | --- | --- | --- | --- | --- | --- |
| AAA | 19832 | 5786 | 5679 | 5768 | 1975 | 1944 | 1944 | 1986 | 1899 | 1952 | 1954 | 1935 | 1876 |
| AAC | 6246 | 1709 | 1707 | 1643 | 550 | 560 | 567 | 587 | 543 | 557 | 504 | 531 | 534 |
| AAG | 7087 | 1859 | 1783 | 1940 | 607 | 619 | 651 | 607 | 630 | 611 | 615 | 679 | 649 |
| AAT | 15037 | 5320 | 5352 | 5428 | 1784 | 1685 | 1769 | 1770 | 1758 | 1743 | 1757 | 1775 | 1742 |
| ACA | 5049 | 1527 | 1564 | 1635 | 542 | 513 | 546 | 492 | 566 | 521 | 537 | 517 | 519 |
| ACC | 2363 | 747 | 755 | 747 | 233 | 253 | 266 | 241 | 273 | 253 | 231 | 203 | 236 |
| ACG | 1029 | 660 | 663 | 684 | 214 | 205 | 197 | 175 | 210 | 208 | 228 | 203 | 203 |
| ACT | 4714 | 1713 | 1745 | 1702 | 526 | 548 | 553 | 536 | 568 | 544 | 552 | 537 | 506 |
| AGA | 5272 | 1784 | 1688 | 1737 | 657 | 635 | 640 | 631 | 600 | 598 | 691 | 644 | 638 |
| AGC | 2754 | 590 | 623 | 586 | 226 | 216 | 200 | 220 | 199 | 181 | 198 | 223 | 204 |
| AGG | 2150 | 973 | 880 | 912 | 293 | 294 | 322 | 292 | 259 | 287 | 314 | 288 | 307 |
| AGT | 4713 | 1700 | 1704 | 1820 | 530 | 595 | 543 | 513 | 492 | 523 | 562 | 549 | 543 |
| ATA | 11126 | 4952 | 5051 | 4886 | 1688 | 1629 | 1637 | 1671 | 1655 | 1659 | 1706 | 1635 | 1606 |
| ATC | 5250 | 1619 | 1570 | 1568 | 507 | 537 | 511 | 518 | 527 | 583 | 531 | 522 | 549 |
| ATG | 5499 | 1450 | 1488 | 1505 | 494 | 547 | 582 | 510 | 529 | 538 | 551 | 514 | 539 |
| ATT | 15079 | 5397 | 5390 | 5419 | 1797 | 1802 | 1835 | 1816 | 1832 | 1799 | 1857 | 1745 | 1726 |

| | | | | | | | | | | | | |
|---|---|---|---|---|---|---|---|---|---|---|---|---|
| CAA | 7427 | 1620 | 1615 | 1661 | 526 | 538 | 568 | 565 | 502 | 508 | 540 | 561 | 529 |
| CAC | 1872 | 700 | 746 | 733 | 225 | 236 | 201 | 255 | 262 | 243 | 222 | 238 | 224 |
| CAG | 2105 | 553 | 653 | 605 | 212 | 206 | 211 | 205 | 203 | 221 | 197 | 203 | 249 |
| CAT | 5375 | 1473 | 1497 | 1388 | 547 | 546 | 519 | 548 | 497 | 539 | 543 | 507 | 528 |
| CCA | 2750 | 727 | 712 | 664 | 235 | 248 | 252 | 267 | 252 | 246 | 218 | 255 | 238 |
| CCC | 569 | 522 | 622 | 513 | 181 | 166 | 170 | 188 | 198 | 183 | 164 | 196 | 174 |
| CCG | 681 | 336 | 390 | 326 | 81 | 113 | 84 | 105 | 109 | 118 | 94 | 109 | 108 |
| CCT | 2181 | 887 | 945 | 885 | 292 | 277 | 293 | 323 | 302 | 307 | 281 | 292 | 286 |
| CGA | 1171 | 635 | 607 | 600 | 189 | 208 | 178 | 196 | 207 | 194 | 204 | 208 | 200 |
| CGC | 508 | 321 | 341 | 319 | 106 | 113 | 89 | 104 | 101 | 107 | 109 | 116 | 94 |
| CGG | 693 | 402 | 365 | 366 | 124 | 93 | 102 | 121 | 137 | 105 | 85 | 97 | 109 |
| CGT | 989 | 671 | 663 | 684 | 214 | 175 | 194 | 199 | 189 | 215 | 164 | 204 | 190 |
| CTA | 4326 | 1664 | 1659 | 1562 | 545 | 515 | 520 | 526 | 532 | 542 | 543 | 524 | 523 |
| CTC | 1786 | 832 | 893 | 841 | 310 | 306 | 295 | 289 | 306 | 316 | 283 | 312 | 279 |
| CTG | 2115 | 577 | 647 | 636 | 215 | 236 | 206 | 207 | 172 | 211 | 196 | 173 | 222 |
| CTT | 6917 | 1950 | 1913 | 1785 | 635 | 646 | 641 | 684 | 653 | 577 | 617 | 614 | 632 |
| GAA | 7190 | 1823 | 1801 | 1812 | 651 | 689 | 675 | 640 | 610 | 655 | 602 | 654 | 660 |
| GAC | 1404 | 585 | 598 | 611 | 208 | 204 | 178 | 195 | 223 | 208 | 189 | 201 | 185 |
| GAG | 1820 | 932 | 833 | 930 | 289 | 289 | 274 | 284 | 275 | 295 | 291 | 291 | 315 |
| GAT | 5225 | 1664 | 1563 | 1555 | 555 | 530 | 523 | 488 | 507 | 534 | 577 | 557 | 564 |
| GCA | 2974 | 572 | 580 | 631 | 228 | 195 | 197 | 241 | 210 | 196 | 196 | 198 | 227 |
| GCC | 710 | 276 | 353 | 288 | 97 | 91 | 110 | 102 | 92 | 118 | 106 | 99 | 106 |
| GCG | 497 | 377 | 321 | 361 | 113 | 97 | 121 | 109 | 108 | 109 | 103 | 104 | 102 |
| GCT | 2973 | 622 | 582 | 585 | 227 | 233 | 218 | 181 | 227 | 192 | 207 | 178 | 198 |
| GGA | 2330 | 958 | 875 | 888 | 283 | 302 | 296 | 297 | 272 | 316 | 308 | 318 | 317 |
| GGC | 676 | 286 | 275 | 283 | 115 | 110 | 100 | 112 | 112 | 108 | 93 | 104 | 97 |
| GGG | 616 | 551 | 546 | 555 | 185 | 200 | 193 | 187 | 170 | 200 | 215 | 174 | 209 |
| GGT | 2446 | 733 | 755 | 728 | 246 | 240 | 281 | 232 | 249 | 246 | 229 | 243 | 227 |
| GTA | 3636 | 1680 | 1606 | 1709 | 516 | 537 | 532 | 518 | 544 | 501 | 513 | 546 | 555 |
| GTC | 1374 | 601 | 577 | 578 | 199 | 198 | 218 | 194 | 217 | 226 | 202 | 203 | 190 |
| GTG | 2083 | 711 | 720 | 777 | 265 | 257 | 238 | 219 | 226 | 237 | 244 | 239 | 243 |
| GTT | 6587 | 1694 | 1736 | 1770 | 575 | 541 | 553 | 560 | 559 | 530 | 551 | 577 | 574 |
| TAA | 13334 | 5401 | 5418 | 5430 | 1732 | 1708 | 1678 | 1693 | 1760 | 1717 | 1762 | 1723 | 1839 |
| TAC | 3369 | 1625 | 1685 | 1624 | 520 | 528 | 509 | 571 | 571 | 535 | 492 | 554 | 501 |
| TAG | 4368 | 1685 | 1596 | 1641 | 498 | 538 | 556 | 475 | 505 | 561 | 587 | 544 | 572 |
| TAT | 11019 | 4895 | 4950 | 4973 | 1635 | 1730 | 1667 | 1648 | 1672 | 1649 | 1711 | 1695 | 1767 |
| TCA | 6993 | 1542 | 1679 | 1586 | 549 | 514 | 546 | 542 | 559 | 575 | 573 | 542 | 570 |

| | | | | | | | | | | | | |
|---|---|---|---|---|---|---|---|---|---|---|---|---|
| TCC | 2302 | 872 | 904 | 854 | 291 | 276 | 293 | 340 | 353 | 277 | 295 | 316 | 283 |
| TCG | 1240 | 681 | 662 | 669 | 183 | 201 | 204 | 193 | 204 | 206 | 220 | 212 | 214 |
| TCT | 5710 | 1794 | 1866 | 1798 | 639 | 645 | 600 | 646 | 634 | 658 | 647 | 669 | 632 |
| TGA | 6550 | 1642 | 1533 | 1602 | 577 | 570 | 560 | 530 | 557 | 525 | 568 | 556 | 590 |
| TGC | 2790 | 616 | 610 | 611 | 190 | 233 | 201 | 219 | 219 | 213 | 205 | 191 | 194 |
| TGG | 2658 | 720 | 689 | 796 | 222 | 285 | 245 | 246 | 241 | 240 | 239 | 293 | 242 |
| TGT | 5123 | 1727 | 1583 | 1619 | 585 | 568 | 547 | 565 | 592 | 546 | 531 | 536 | 566 |
| TTA | 13519 | 5470 | 5525 | 5585 | 1755 | 1753 | 1760 | 1765 | 1793 | 1802 | 1733 | 1777 | 1758 |
| TTC | 7131 | 1839 | 1864 | 1809 | 644 | 631 | 659 | 669 | 660 | 672 | 632 | 680 | 571 |
| TTG | 7918 | 1700 | 1745 | 1689 | 576 | 579 | 583 | 563 | 562 | 559 | 541 | 553 | 612 |
| TTT | 20219 | 5886 | 5876 | 5921 | 1997 | 1929 | 2004 | 2034 | 1960 | 2010 | 1995 | 1969 | 1963 |

Figure 20. The table of frequencies of 64 triplets in long nucleotide sequences $S_0$, $S_{1/0}$, $S_{1/1}$, $S_{1/2}$, $S_{2/00}$, $S_{2/01}$, $S_{2/02}$, $S_{2/10}$, $S_{2/11}$, $S_{2/12}$, $S_{2/20}$, $S_{2/21}$, $S_{2/22}$ at the first three levels of FGN-3 of Mycoplasma crocodyli MP145 chromosome, complete genome (NCBI Reference Sequence: NC_014014.1 http://www.ncbi.nlm.nih.gov/nuccore/294155300).

**The Symmetry Principle № 3**: for each of long nucleotide sequences at each level of FGN-3 the following rules hold true: sum of the frequencies of all the oligonucleotides, that begin with the letter A, approximately equal to the sum of the frequencies of all the oligonucleotides that begin with the letter T; sum of the frequencies of all the oligonucleotides, that begin with the letter C, approximately equal to the sum of the frequencies of all the oligonucleotides that begin with the letter T.

In particularly, these rules hold true not only for long sequences at lower levels of FGN-3 but also for an initial long sequence $S_0$. Figure 21 illustrates the Symmetry Principle № 3 by examples of n-plets (n=2, 3, 4, 5) in sequences $S_0$, $S_{1/0}$,…, $S_{2/22}$ of the first levels in FGN-3 of the same genome as on Figures 18-20.

The total frequencies of the sets of duplets in sequences of FGN-3:

| | $S_0$ | $S_{1/0}$ | $S_{1/1}$ | $S_{1/2}$ | $S_{2/00}$ | $S_{2/01}$ | $S_{2/02}$ | $S_{2/10}$ | $S_{2/11}$ | $S_{2/12}$ | $S_{2/20}$ | $S_{2/21}$ | $S_{2/22}$ |
|---|---|---|---|---|---|---|---|---|---|---|---|---|---|
| F(A) | 169757 | 56674 | 56197 | 56747 | 19065 | 18857 | 18836 | 18764 | 18756 | 18713 | 18949 | 18815 | 18871 |
| F(T) | 171420 | 57022 | 57247 | 57179 | 18870 | 18937 | 19067 | 19102 | 19155 | 19040 | 19161 | 19105 | 19083 |
| F(C) | 62531 | 20763 | 21486 | 20600 | 6929 | 6903 | 6882 | 7113 | 7148 | 7178 | 6772 | 6900 | 6800 |
| F(G) | 63471 | 21265 | 20794 | 21198 | 7044 | 7211 | 7123 | 6929 | 6849 | 6977 | 7026 | 7088 | 7154 |

The total frequencies of the sets of triplets in sequences of FGN-3:

| | $S_0$ | $S_{1/0}$ | $S_{1/1}$ | $S_{1/2}$ | $S_{2/00}$ | $S_{2/01}$ | $S_{2/02}$ | $S_{2/10}$ | $S_{2/11}$ | $S_{2/12}$ | $S_{2/20}$ | $S_{2/21}$ | $S_{2/22}$ |
|---|---|---|---|---|---|---|---|---|---|---|---|---|---|
| F(A) | 113200 | 37786 | 37642 | 37980 | 12623 | 12582 | 12763 | 12565 | 12540 | 12557 | 12788 | 12500 | 12377 |
| F(T) | 114243 | 38095 | 38185 | 38207 | 12593 | 12688 | 12612 | 12699 | 12842 | 12745 | 12731 | 12810 | 12874 |
| F(C) | 41465 | 13870 | 14268 | 13568 | 4637 | 4622 | 4523 | 4782 | 4622 | 4632 | 4460 | 4609 | 4585 |
| F(G) | 42541 | 14065 | 13721 | 14061 | 4752 | 4713 | 4707 | 4559 | 4601 | 4671 | 4626 | 4686 | 4769 |

The total frequencies of the sets of 4-plets in sequences of FGN-3:

| | $S_0$ | $S_{1/0}$ | $S_{1/1}$ | $S_{1/2}$ | $S_{2/00}$ | $S_{2/01}$ | $S_{2/02}$ | $S_{2/10}$ | $S_{2/11}$ | $S_{2/12}$ | $S_{2/20}$ | $S_{2/21}$ | $S_{2/22}$ |
|---|---|---|---|---|---|---|---|---|---|---|---|---|---|
| F(A) | 84955 | 28475 | 27999 | 28493 | 9522 | 9391 | 9274 | 9342 | 9417 | 9434 | 9453 | 9379 | 9371 |
| F(T) | 85573 | 28439 | 28601 | 28639 | 9531 | 9532 | 9624 | 9552 | 9570 | 9538 | 9663 | 9573 | 9555 |
| F(C) | 31189 | 10286 | 10758 | 10270 | 3409 | 3438 | 3499 | 3573 | 3578 | 3594 | 3331 | 3423 | 3356 |
| F(G) | 31867 | 10662 | 10504 | 10460 | 3492 | 3593 | 3557 | 3487 | 3389 | 3388 | 3507 | 3579 | 3672 |

The total frequencies of the sets of 5-plets in sequences of FGN-3:

| | $S_0$ | $S_{1/0}$ | $S_{1/1}$ | $S_{1/2}$ | $S_{2/00}$ | $S_{2/01}$ | $S_{2/02}$ | $S_{2/10}$ | $S_{2/11}$ | $S_{2/12}$ | $S_{2/20}$ | $S_{2/21}$ | $S_{2/22}$ |
|---|---|---|---|---|---|---|---|---|---|---|---|---|---|
| F(A) | 67729 | 22626 | 22512 | 22788 | 7551 | 7506 | 7542 | 7491 | 7417 | 7431 | 7587 | 7445 | 7539 |
| F(T) | 68688 | 22951 | 22918 | 22764 | 7620 | 7613 | 7684 | 7626 | 7668 | 7677 | 7619 | 7700 | 7534 |

| F(C) | 25144 | 8242 | 8503 | 8217 | 2780 | 2762 | 2701 | 2851 | 2907 | 2849 | 2704 | 2785 | 2736 |
|---|---|---|---|---|---|---|---|---|---|---|---|---|---|
| F(G) | 25304 | 8470 | 8356 | 8520 | 2812 | 2882 | 2836 | 2795 | 2771 | 2806 | 2853 | 2833 | 2954 |

Figure 21. The illustration of the Symmetry Principle № 3 in the case of Mycoplasma crocodyli MP145 chromosome, complete genome (NCBI Reference Sequence: NC_014014.1 http://www.ncbi.nlm.nih.gov/nuccore/294155300). Here F(A), F(T), F(C) and F(G) mean sum of the frequencies of oligonucleotides (or n-plets) that begin with the letters A, T, C or G correspondingly. The tables show the F(A) ≈ F(T) and F(C) ≈ F(G) for sets of n-plets (n=2, 3, 4, 5) in each of long nucleotide sequences $S_0$, $S_{1/0}$, $S_{1/1}$, $S_{1/2}$, $S_{2/00}$, $S_{2/01}$, $S_{2/02}$, $S_{2/10}$, $S_{2/11}$, $S_{2/12}$, $S_{2/20}$, $S_{2/21}$, $S_{2/22}$ at the first three levels of FGN-3 of this genome.

One can note that each of 4 quadrants of the genomatrices [C T; A G]$^{(n)}$ (Figure 1) contain all oligonucleotides that begin with one of these 4 letters C, T, A or G. In these genomatrices each oligonucleotide and its complementary oligonucleotide are disposed inverse-symmetrical relative to the centre of the appropriate matrix. In accordance with the Symmetry Principle № 3, the total frequencies of oligonucleotides in both quadrants along the main diagonal of these gemomatrices are approximately equal each other (F(C) ≈ F(G)); the total frequencies of oligonucleotides in both quadrants along the second diagonal of these gemomatrices are also approximately equal each other (F(A) ≈ F(T)).

An additional illustration of the Symmetry Principle № 3 is obtained from data about the whole human genome from the work [Perez, 2010]. This genome contains 2.843.411.612 triplets. Figure 22 shows the total frequencies of sets of triplets that begin with one of the four letters A, C, G or T.

| The total frequencies of the sets of triplets that begin with A | The total frequencies of the sets of triplets that begin with T |
|---|---|
| f(ACC+ACT+ACA+ACG+ ATC+ATT+ATA+ATG+ AAC+AAT+AAA+AAG+ AGC+AGT+AGA+AGG)= **839.827.642** | f(TGG+TGA+TGT+TGC+ TAG+TAA+TAT+TAC+ TTG+TTA+TTT+TTC+ TCG+TCA+TCT+TCC)= **841.214.589** |

| The total frequencies of the sets of triplets that begin with C: | The total frequencies of the sets of triplets that begin with G: |
|---|---|
| f(CCC+CCT+CCA+CCG+ CTC+CTT+CTA+CTG+ CAC+CAT+CAA+CAG+ CGC+CGT+CGA+CGG) = **581.026.275** | f(GGG+GGA+GGT+GGC+ GAG+GAA+GAT+GAC+ GCG+GTA+GTT+GTC+ GTG+GCA+GCT+GCC)= **581.343.106** |

Figure 22. The approximate equality of the total frequencies of sets of triplets that begin with letters A and T (upper table) and with letters C and G (bottom table) in the case of the whole human genome. Initial data about frequencies of separate triplets are taken from the work [Perez, 2010].

The Symmetry Principle № 3 speaks about pairs of complete sets of complementary olygonucleotides that begin with complementary letters (A=T and C=G). If one consider sub-sets of complementary oligonucleotides from these complete sets, the equality of their total frequences is violated, and this violation is the more the less scale of sub-sets is under consideration. For example, in the case of FGN-3 of long nucleotide sequences, the equality of the total frequencies of two sets of 4 pentaplets, which are disposed in two complementary (2*2)-sub-sub-quadrants of the matrix [C T; A G]$^{(5)}$, is performed less accurately than the equality of the total frequencies of two sets of 16 pentaplets in two complementary (4*4)-sub-quadrants in this matrix.

The author would like to mention about the additional Symmetry Principles № 4, which can be interpreted as simple consequences of the generalized Symmetry Principle № 1 and the Symmetry Principle № 2, though today it is difficult to say what of these Principles are more fundamental. Our article [Petoukhov, 2012a, http://arxiv.org/abs/1102.3596] describes the method of dyadic-shift numeration of triplets inside genomatrices [C A; G T]$^{(3)}$ and [C T; G A]$^{(3)}$ on the base of the binary-oppositional attributes of nucleotides A, C, G, T. In the result, the sets of 8 triplets with identical dyadic-shift numerations 000, 001,…., 111 arise, and they obey the following rule in long nucleotide sequences: the frequency of triplets with a dyadic shift numeration (for example "010") is approximately equal to the frequency of the triplet with inverse dyadic shift numeration ("101" in this example) (other oligonucleotides can be dyadic-shift numerated by analogy). Inverse numeration is produced by replacement of numbers 0➔1 and 1➔0. Such sets of oligonucleotides are main participants in the **Symmetry Principle № 4**:

- for each of long nucleotide sequences at each level of FGN-3, the total frequency of each set of oligonucleotides, which have the same dyadic-shift numeration on the base of the method of such numeration from the work [Petoukhov, 2012a], is equal approximately to the total frequency of the set of oligonucleotides, which have inverse dyadic-shift numeration.

Now let us represent the Symmetry Principle № 5 which speaks about reading frame shifts, deletion mutations and also positional permutations in oligonucleotides. Concerning those DNA-sequences (including the mentioned genome on Figures 18-21), which have been tested till today in the author's laboratory, we have revealed the following phenomenological facts (this study is continued now for a wide list of DNA-sequences of different organisms and organelles):

- a transformation of long nucleotide sequences by means of a reading frame shift in them preserves implementations of all described Symmetry Principles inside new long nucleotide sequences (in our tests, a reading frame shift means that the reading of sequence does not begin with its first position, but with one of subsequent positions; the missing fragment of the sequence can be moved into the end of the sequence, and in this case a reading frame shift leads to a simple change of order of all sequences at each of lower levels of FGN);
- a transformation of long nucleotide sequences by means of a deletion mutation (when their short parts are missing) preserves implementations of all described Symmetry Principles in new long nucleotide sequences.

One should consider separately the question about positional permutations in oligonucleotides. The theory of noise-immunity coding pays a special attention to permutations of elements of transmitted signals. It is obvious that for different n-plets different quantities of variants of permutation of their positions exist:

- for duplets two variants of positional permutations exist (1-2 and 2-1);
- for triplets six variants of positional permutations exist (1-2-3, 2-3-1, 3-1-2, 3-2-1, 2-1-3, 1-3-2);
- for 4-plets 24 variants of positional permutations exist (1-2-3-4, 2-3-4-1, …..);
- for 5-plets 120 variants of positional permutations exist (1-2-3-4-5, 2-3-4-5-1, …..).

It is also obvious that if a long nucleotide sequence is interpreted as a sequence of a certain type of oligonucleotides (duplets, or triplets, or 4-plets, or 5-plets, …), and one of possible positional permutations is done simultaneously inside all of its oligonucleotides, then a quite new long nucleotide sequence appears. For example if we have initially a sequence of triplets CGA-TAA-AGC-GTC-TAG-CGC-ATC -…, then after changing of the positional order from the initial order 1-2-3 to new order 2-3-1 inside each of triplets, we obtain the quite different sequence GAC-AAT-GCA-TCG-AGT-GCC-TCA -… . But FGN-3 for this new long nucleotide sequence is obeyed the same Symmetry Principles №№ 1-3 which are described above. We name simultaneous positional permutations inside all oligonucleotides of a certain type as "collective positional permutations" inside these oligonucleotides. The author proposes a brief formulation of these phenomenological facts by the following way.

**The Symmetry Principle № 5**:
- reading frame shifts and deletion mutations in long nucleotide sequences and also collective positional permutations inside their oligonucleotides don't essentially violate implementations of all the Symmetry Principles described in this article for long nucleotide sequences and their fractal genetic net (FGN-3).

It appears that the described FGN-3 and fractal-like properties of long genetic sequences, which are related to the invariance of these Symmetry Principles, have a biological value (a biological sense) associated with mutational changes of such sequences and with evolutionary producing new types of DNA-sequences. The author supposes that mechanisms of biological evolution use these permutational and other described properties of long nucleotide sequences in producing new biological organisms and organelles. For instance new DNA-sequences can be constructed in the course of biological evolution of organisms by means of combinatorics of nucleotide sequences from different levels of FGN (including genetic crossing among long nucleotide sequences from different levels of FGN by analogy with well-known examples of genetic crossing). One should note here that the question about permutational properties of DNA-sequences is very important because some biological organisms differ each from other only by permutations in their DNA-sequences (see for example the book [Pevzner, 2000]). The proposed method of FGN is the new effective and useful method in the field of bioinformatics, molecular genetics and evolutionary biology.

The author supposes that for long nucleotide sequences the following Symmetry Principle № 6 exists also, which compares a total frequency $F_{even}$ of all n-plets (oligonucleotides) in columns with even numerations in matrices $[C\ T;\ A\ G]^{(n)}$ (n=1, 2, 3, …) with a total frequency $F_{odd}$ of all n-plets in columns with odd numerations in the same matrices. Let us numerate columns from left to right in these matrices by numbers 0, 1, 2, 3, … and then let us pay our attention separately to the first set of n-plets in columns with even numeration (0, 2, 4, …) and to the second set of n-plets in columns with odd numeration. We name conditionally the first set as the set of the even type (or briefly, the even-set) and the second set as the set of the odd type (or the odd-set). For example, the genomatrix $[C\ T;\ A\ G]^{(2)}$ contains two sets of duplets (Figure 11): 1) the even-set, which represents the columns with even numerations 0 and 2, contains the 8 duplets (CC, CA, AC, AA, TC, TA, GC, GA); 2) the odd-set, which represents the columns with odd numerations 1 and 3, contains the 8 duplets (CT, CG, AT, AG, TT, TG, GT, GG). Or in the genomatrix $[C\ T;\ A\ G]^{(3)}$ we have two sets of triplets (Figure 11): 1) the even-set, which represents the columns with even numerations 0, 2, 4, 6, contains the 32 triplets (CCC, CCA, CAC, CAA, ACC, ACA, AAC, AAA, CTC, CTA, CGC, CGA, ATC, ATA, AGC, AGA, TCC, TCA, TAC, TAA, GCC, GCA, GAC, GAA, TTC, TTA, TGC, TGA, GTC, GTA, GGC, GGA); 2) the odd-set, which represents the columns with odd numerations 1, 3, 5, 7, contains the 32 triplets (CCT, CCG, CAT, CAG, ACT, ACG, AAT, AAG, CTT, CTG, CGT, CGG, ATT, ATG, AGT, AGG, TCT, TCG, TAT, TAG, GCT, GCG, GAT, GAG, TTT, TTG, TGT, TGG, GTT, GTG, GGT, GGG).

We are studying of long nucleotide sequences to compare the total frequencies $F_{even}$ and $F_{odd}$ of n-plets from the even-sets and from the odd-sets at different levels of FGN-3. Our initial research results confirm that these frequencies are approximately equal. Figure 23 confirms this fact for the case of duplets from the even-set and the odd-set of sequences $S_0$, $S_{1/0}$, $S_{1/1}$, $S_{1/2}$, $S_{2/00}$, …, $S_{3/222}$ at the first four levels of the FGN-3 of the above-considered Mycoplasma crocodyli MP145 chromosome, complete genome (NCBI Reference Sequence: NC_014014.1 http://www.ncbi.nlm.nih.gov/nuccore/294155300). More precisely, Figure 23 shows percentage differences between the total frequencies $F_{even}$ and $F_{odd}$ of duplets from the even-set and the odd-set of the Mycoplasma crocodyli MP145 chromosome. Here $F_{odd}$ is taken as 100%, and its percentage difference $\Delta$ in relation to $F_{even}$ is calculated in each case by means of the formula $\Delta = 100*(1 - F_{evev}/F_{odd})$. For example, in the initial nucleotide sequence $S_0$, duplets from the even-set are met 232415 times ($F_{even}$=232415) and duplets from the odd-set are met 234764 times ($F_{odd}$=234764). In accordance with the formula we have $\Delta S_0 = 100*(1 - 232415/234764) = 1,001\%$ (see $\Delta S_0$ in Figure 23).

| $\Delta S_0$ | $\Delta S_{1/0}$ | $\Delta S_{1/1}$ | $\Delta S_{1/2}$ | $\Delta S_{2/0}$ | $\Delta S_{2/01}$ | $\Delta S_{2/02}$ | $\Delta S_{2/10}$ |
|---|---|---|---|---|---|---|---|
| 1,001% | 1,618% | 0,259% | 1,598% | 2,225% | 1,749% | 1,142% | -0,611% |

| $\Delta S_{2/11}$ | $\Delta S_{2/12}$ | $\Delta S_{2/20}$ | $\Delta S_{2/21}$ | $\Delta S_{2/22}$ | $\Delta S_{3/000}$ | $\Delta S_{3/001}$ | $\Delta S_{3/002}$ |
|---|---|---|---|---|---|---|---|
| 0,300% | 0,997% | 0,974% | -0,008% | 1,991% | 0,737% | 0,438% | 2,781% |

| $\Delta S_{3/010}$ | $\Delta S_{3/011}$ | $\Delta S_{3/012}$ | $\Delta S_{3/020}$ | $\Delta S_{3/021}$ | $\Delta S_{3/022}$ | $\Delta S_{3/100}$ | $\Delta S_{3/101}$ |
|---|---|---|---|---|---|---|---|
| 1,446% | 1,605% | 1,332% | 0,553% | 1,309% | 0,461% | -3,362% | 0,415% |

| $\Delta S_{3/102}$ | $\Delta S_{3/110}$ | $\Delta S_{3/111}$ | $\Delta S_{3/112}$ | $\Delta S_{3/120}$ | $\Delta S_{3/121}$ | $\Delta S_{3/122}$ | $\Delta S_{3/200}$ |
|---|---|---|---|---|---|---|---|
| -0,510% | 0,783% | 0,392% | 0,046% | 1,423% | 0,966% | 1,058% | 1,241% |

| $\Delta S_{3/201}$ | $\Delta S_{3/202}$ | $\Delta S_{3/210}$ | $\Delta S_{3/211}$ | $\Delta S_{3/212}$ | $\Delta S_{3/220}$ | $\Delta S_{3/221}$ | $\Delta S_{3/222}$ |
|---|---|---|---|---|---|---|---|
| 3,3640% | 2,601% | 0,507% | 1,149% | 0,069% | 4,7188% | 0,966% | 2,894% |

Figure 23. Percentage differences $\Delta$ between the total frequencies $F_{even}$ and $F_{odd}$ of duplets from the even-set of duplets (CC, CA, AC, AA, TC, TA, GC, GA) and from the odd-set of duplets (CT, CG, AT, AG, TT, TG, GT, GG) in sequences $S_0$, $S_{1/0}$, $S_{1/1}$, $S_{1/2}$, $S_{2/00}$, …, $S_{3/222}$ at the first four levels of the FGN-3 of the Mycoplasma crocodyli MP145 chromosome, complete genome (NCBI Reference Sequence: NC_014014.1 http://www.ncbi.nlm.nih.gov/nuccore/294155300). The initial sequence $S_0$ contains 934379 nucleotides. Percentages are shown with a precision up to three decimal places.

Data of Figure 24 confirms the approximate equality of $F_{even}$ and $F_{odd}$ of triplets in the same sequences $S_0$, $S_{1/0}$, $S_{1/1}$, $S_{1/2}$, $S_{2/00}$, …, $S_{3/222}$ at the first four levels of the FGN-3 of the Mycoplasma crocodyli MP145 chromosome.

| $\Delta S_0$ | $\Delta S_{1/0}$ | $\Delta S_{1/1}$ | $\Delta S_{1/2}$ | $\Delta S_{2/0}$ | $\Delta S_{2/01}$ | $\Delta S_{2/02}$ | $\Delta S_{2/10}$ |
|---|---|---|---|---|---|---|---|
| 1,455% | 1,472% | 0,741% | 2,074% | 0,605% | 1,679% | 2,359% | -0,713% |

| $\Delta S_{2/11}$ | $\Delta S_{2/12}$ | $\Delta S_{2/20}$ | $\Delta S_{2/21}$ | $\Delta S_{2/22}$ | $\Delta S_{3/000}$ | $\Delta S_{3/001}$ | $\Delta S_{3/002}$ |
|---|---|---|---|---|---|---|---|
| -0,830% | 0,179% | 2,268% | 0,237% | 2,697% | -1,837% | -0,644% | -3,296% |

| $\Delta S_{3/010}$ | $\Delta S_{3/011}$ | $\Delta S_{3/012}$ | $\Delta S_{3/020}$ | $\Delta S_{3/021}$ | $\Delta S_{3/022}$ | $\Delta S_{3/100}$ | $\Delta S_{3/101}$ |
|---|---|---|---|---|---|---|---|
| -0,713% | 0,260% | 3,459% | 1,804% | 1,224% | 3,459% | -3,260% | -1,555% |

| $\Delta S_{3/102}$ | $\Delta S_{3/110}$ | $\Delta S_{3/111}$ | $\Delta S_{3/112}$ | $\Delta S_{3/120}$ | $\Delta S_{3/121}$ | $\Delta S_{3/122}$ | $\Delta S_{3/200}$ |
|---|---|---|---|---|---|---|---|
| -3,475% | 3,358% | -2,156% | -1,485% | -0,017% | 0,536% | -1,379% | 0,017% |

| $\Delta S_{3/201}$ | $\Delta S_{3/202}$ | $\Delta S_{3/210}$ | $\Delta S_{3/211}$ | $\Delta S_{3/212}$ | $\Delta S_{3/220}$ | $\Delta S_{3/221}$ | $\Delta S_{3/222}$ |
|---|---|---|---|---|---|---|---|
| 1,018% | 0,812% | 4,492% | -1,239% | 1,052% | 2,686% | -1,168% | 2,212% |

Figure 24. Percentage differences $\Delta$ between the total frequencies $F_{even}$ and $F_{odd}$ of triplets from the even-set of triplets (CCC, CCA, CAC, CAA, ACC, ACA, AAC, AAA, CTC, CTA, CGC, CGA, ATC, ATA, AGC, AGA, TCC, TCA, TAC, TAA, GCC, GCA, GAC, GAA, TTC, TTA, TGC, TGA, GTC, GTA, GGC, GGA) and from the odd-set of triplets (CCT, CCG, CAT, CAG, ACT, ACG, AAT, AAG, CTT, CTG, CGT, CGG, ATT, ATG, AGT, AGG, TCT, TCG, TAT, TAG, GCT, GCG, GAT, GAG, TTT, TTG, TGT, TGG, GTT, GTG, GGT, GGG) in sequences $S_0$, $S_{1/0}$, $S_{1/1}$, $S_{1/2}$, $S_{2/00}$, …, $S_{3/222}$ at the first four levels of the FGN-3 of the Mycoplasma crocodyli MP145 chromosome, complete genome (NCBI Reference Sequence: NC_014014.1 http://www.ncbi.nlm.nih.gov/nuccore/294155300). The initial sequence $S_0$ contains 934379 nucleotides. Percentages are shown with a precision up to three decimal places.

In a favor of existence of the Symmetry Principle № 6, the additional evidence is the following. Let us compare the total frequencies $F_{even}$ and $F_{odd}$ of triplets from the whole human genome, which contains the huge number 2.843.411.612 (about three billion) triplets.

Figure 25 shows frequencies of each of 64 triplets in the whole human genome from the article [Perez, 2010]. One can see from this Figure that frequencies of different triplets can differ in many times. For example, the frequency of the triplet CGA is equal to 6.251.611 and the frequency of the triplet TTT is equal to 109.591.342. These two frequencies differ in 18 times approximately. But values of the total frequencies $F_{even}$ and $F_{odd}$ of triplets are equal to within 0.12%.

| triplet | triplet frequency | triplet | triplet frequency | triplet | triplet frequency | triplet | triplet frequency |
|---|---|---|---|---|---|---|---|
| AAA | 109143641 | CAA | 53776608 | GAA | 56018645 | TAA | 59167883 |
| AAC | 41380831 | CAC | 42634617 | GAC | 26820898 | TAC | 32272009 |
| AAG | 56701727 | CAG | 57544367 | GAG | 47821818 | TAG | 36718434 |
| AAT | 70880610 | CAT | 52236743 | GAT | 37990593 | TAT | 58718182 |
| ACA | 57234565 | CCA | 52352507 | GCA | 40907730 | TCA | 55697529 |
| ACC | 33024323 | CCC | 37290873 | GCC | 33788267 | TCC | 43850042 |
| ACG | 7117535 | CCG | 7815619 | GCG | 6744112 | TCG | 6265386 |
| ACT | 45731927 | CCT | 50494519 | GCT | 39746348 | TCT | 62964984 |
| AGA | 62837294 | CGA | 6251611 | GGA | 43853584 | TGA | 55709222 |
| AGC | 39724813 | CGC | 6737724 | GGC | 33774033 | TGC | 40949883 |
| AGG | 50430220 | CGG | 7815677 | GGG | 37333942 | TGG | 52453369 |
| AGT | 45794017 | CGT | 7137644 | GGT | 33071650 | TGT | 57468177 |
| ATA | 58649060 | CTA | 36671812 | GTA | 32292235 | TTA | 59263408 |
| ATC | 37952376 | CTC | 47838959 | GTC | 26866216 | TTC | 56120623 |
| ATG | 52222957 | CTG | 57598215 | GTG | 42755364 | TTG | 54004116 |
| ATT | 71001746 | CTT | 56828780 | GTT | 41557671 | TTT | 109591342 |

Figure 25. Quantities of repetitions of each triplet in the whole human genome (data are taken from the work [Perez, 2010])

Really from Figure 25, one can calculate the total frequencies $F_{even}$ and $F_{odd}$ of triplets from the whole human genome and receive the confirmation of their approximate equality:
- $F_{even}$ = **1.420.853.821** for the even-set of 32 triplets (CCC, CCA, CAC, CAA, ACC, ACA, AAC, AAA, CTC, CTA, CGC, CGA, ATC, ATA, AGC, AGA, TCC, TCA, TAC, TAA, GCC, GCA, GAC, GAA, TTC, TTA, TGC, TGA, GTC, GTA, GGC, GGA);
- $F_{odd}$ = **1.422.557.791** for the odd-set of 32 triplets (CCT, CCG, CAT, CAG, ACT, ACG, AAT, AAG, CTT, CTG, CGT, CGG, ATT, ATG, AGT, AGG, TCT, TCG, TAT, TAG, GCT, GCG, GAT, GAG, TTT, TTG, TGT, TGG, GTT, GTG, GGT, GGG);
- The percentage difference between these $F_{even}$ and $F_{odd}$ is equal to 0,12%.

The appropriate **Symmetry Principle № 6**, which should be studied in additional researches of long nucleotide sequences, can be formulated in the following manner:

- for each of long nucleotide sequences at each level of FGN-3 the following rules hold true: the total frequency $F_{even}$ of all n-plets from the columns with even numerations inside the genomatrix [C T; A G]$^{(n)}$ (here n =1, 2, 3, at least) is approximately equal to the total frequency $F_{odd}$ of all n-plets from the columns with odd numerations inside the same genomatrix.

Our approaches concern to one of important questions of modern science also: the existence of fractal images in genetic systems. A number of publications are devoted to fractal features of genetic texts [Jeffry, 1990; Pellionisz et al, 2011; Petoukhov, 2008b; Petoukhov, He, 2009; Skaletsky et al., 2003; Yam, 1995, etc]. Interesting data about fractal approaches in genetics, including materials about an important connection of fractal defects with cancer, are presented at the WEB site by A.Pellionisz [http://www.junkdna.com/the_genome_is_fractal.html]). Researches in this direction proceed all over the world. In this article the author proposes Fractal Genetics Nets (FGN) as a new tool to study fractal-like properties of long DNA-sequences and also describes new fractal-like properties of such nucleotide sequences. The author supposes that these FGN and fractal-like properties of long nucleotide sequences can lead to new principles and systems in the field of signal processing, recognition of images and artificial intellect. They are used now in the author's laboratory to create new genetic algorithms for different mathematical and technological applications.

We plan to publish in the future other our results on study of FGN and the Symmetry Principles concerning to a wide list of long DNA-sequences of different organells and organisms from different taxonomical classes.

**12 Conclusion remarks**

Many general bioinformation properties exist which should be studied in the field of bioinformatics:
- noise-immunity of genetic coding;
- management and synchronization in a huge hierarchy of cyclic bioprocesses;
- compression of inherited information data;
- primary structures of proteins, etc.

The list of these scientific problems can be added by a list of mathematical and technological problems from the field of genetic algorithms which is developed intensively in world science during last decades (for example see [Goldberg, Korb, Deb, 1989; Forrest, Mitchell,1991]). One can think that results of matrix genetics will be useful in many directions of such studies.

Described results demonstrate that the matrix approach to hierarchical systems of the genetic code can discover hidden interrelations among these systems. The matrix representation of the genetic systems is the effective cognitive form to reveal their algebraic properties, etc. Matrix genetics gives new knowledge about deep analogies between the structure of the genetic code and methods of the theory of digital signal processing. Matrix genetics proposes new mathematical models of biological self-reproduction systems. In our opinion, the condition of noise immunity is the basis of many structural peculiarities of the genetic code systems. The utilization of this condition in matrix analysis of the genetic code systems can lead to discovers of new biological rules and of new mathematical models. It can lead also to new effective decisions in the fields of DNA-computers, quantum computers, nanotechnologies, spectral analysis and theoretical biology. The algebraic essence of the code degeneracy, reflected in mosaics of the described genomatrices, are represented in the publications [Petoukhov, arXiv:0803.3330, arXiv:0805.4692, arXiv:0809.2714; Petoukhov, 2008b; Petoukhov, He, 2009].

The revealed connection of genetic code structures with projector operators is one of new promising ways to introduce powerful algebraic methods into bioinformatics and genetic algorithms. Special attention should be paid to mentioned Boolean algebras of genoprojectors because Boolean algebras are explores in many scientific fields from psychology to computers.
How Boolean algebras of genoprojectors are connected with physiological computers of biological organisms? It is one of many thematic questions which should be studied in future.

The described data about natural possibilities of representations of ensembles of molecular-genetic elements in a form of families of unitary matrices testify that the nature has chosen such scheme of the genetic code which is related with unitary symmetries and with a special set of cyclic groups of unitary transformations. Why the nature has constructed the genetic alphabet which consists of the four nitrogenous bases with their binary-oppositional attributes? The possible reason of this is that the simplest unitary matrix consists of four complex numbers. This thesis is supported by the fact that the genetic molecules belong to the world of quantum mechanics, where unitary transformations play an important role since the evolution of a closed quantum system is unitary.

This version of the article contains new materials about the method of fractal genetic nets (FGN) and about new Symmetry Principles in long nucleotide sequences. These materials represent new tools to study hidden regularities of molecular-genetic systems. The described Symmetry Principles are leading to new understanding and new models of genetic information. In our opinion, materials of this article are important to develop algebraic biology.

**Acknowledgments**: Described researches were made by the author in the frame of a long-term cooperation between Russian and Hungarian Academies of Sciences and in the frame of programs of "International Society of Symmetry in Bioinformatics" (USA, http://polaris.nova.edu/MST/ISSB) and of "International Symmetry Association" (Hungary, http://symmetry.hu/) . The author is grateful to Frolov K.V., Darvas G., Ne'eman Y., He M., D.Pavlov, A.Pellionisz, Smolianinov V.V., Vladimirov Y.S. for their support. Special thanks to V.I.Svirin for his software to analyse long nucleotide sequences by means of fractal genetic nets. Special thanks also to I.V.Stepanyan for his software to analyse some fractal properties of long genetic sequences. Some results of this paper have been obtained thanks to the Russian State scientific contract P377 from July 30, 2009.